\documentclass[final,times,authoryear]{elsarticle}
\makeatletter
\def\ps@pprintTitle{%
 \let\@oddhead\@empty
 \let\@evenhead\@empty
 \def\@oddfoot{\centerline{\thepage}}%
 \let\@evenfoot\@oddfoot}

\makeatother

\usepackage{graphicx,amssymb,amsfonts,amsmath,color,rotating}
\usepackage{subfigure,tablefootnote}
\usepackage{wrapfig}
\usepackage{lscape}
\usepackage{rotating}
\usepackage{setspace}
\usepackage[all]{nowidow}

\begin{document}

\begin{frontmatter}



\title{Interconnectedness in the Global Financial Market}

\author[MR1,MR2]{Matthias Raddant}
\author[JHU]{Dror Y. Kenett}

\address[MR1]{Department of Economics, Kiel University, Olshausenstra{\ss}e 40, 24118 Kiel, Germany}
\address[MR2]{Kiel Institute for the World Economy, Kiellinie 66, 24105 Kiel, Germany}

\address[JHU]{Applied Economics, Zanvyl Krieger School of Arts and Sciences,\\ Johns Hopkins University, Washington, D.C. 20036, USA}
\cortext[cor1]{Corresponding author email:\\ raddant@economics.uni-kiel.de (Matthias Raddant), drorkenett@gmail.com (Dror Y. Kenett)}

\begin{abstract}
The global financial system is highly complex, with cross-border interconnections and interdependencies. In this highly interconnected environment, local financial shocks and events can be easily amplified and turned into global events.  This paper analyzes the dependencies among nearly 4,000 stocks from 15 countries. The stock returns are normalized by the estimated volatility using a GARCH framework, integrated with a robust regression process to estimate pairwise statistically significant relationships between stocks from different countries. The estimation results are used as a measure of statistical interconnectedness, and to derive network representations, both by country and by sector. We find that the Energy, Materials, and Financial sectors play a leading role in connecting markets, and that this role has increased over time for the Energy and Materials sectors. Our results thus confirm the role of global sectoral factors in stock market dependencies. Moreover, our results also show that the dependencies are rather volatile and that heterogeneity among stocks is a non-negligible aspect of this volatility. The transmission mechanism between financial markets is thus not stable, but rather governed by both changes in volatility and changes in the stocks' contributions to the statistical interdependence across countries. 
\end{abstract}


\end{frontmatter}
\vspace{1cm}


\newpage
\doublespacing
\section{Introduction}

The last decades have in general exhibited an increase in political and economic openness, mostly accompanied by an increase in financial integration. Analyzing and managing macroeconomic financial risks have become increasingly important as global markets have become increasingly connected. The sheer size of these markets and the number of products available require models that can provide a hierarchical or topological simplification of this system. 

The financial and economic crisis of the past few decades have shown that contagion can often not be explained by dependencies in the real part of the economy. Distress of financial activity can by itself create contagion which consequences then go beyond the financial sphere.  Numerous studies have therefore looked at dependencies of financial markets, especially if they amplify shocks in times of crisis. The transmission mechanisms in financial markets have been claimed to be relatively stable over time \citep{rigobon_jie,det_inte}, but this claim is debated \citep{corsetti}. Studies that examine determinants of comovement find that structural similarity of the countries' economies explains only partially the level of comovement of their financial markets. This resulted in a debate about the influence of global sectoral factors \citep{dutt,bekhod,segm_equity}. Previous results hint at an increase in the importance of these factors. \cite{forbchinn} find that cross-country factors and global sectoral factors both are important determinants of stock returns. They also note that changes in global linkages over time might make it difficult to disentangle different influences on asset market comovement. 

Despite the size and complexity of global financial markets, most studies have used data sets that compare markets on a rather high level of aggregation, mostly in the form of indices. In order to uncover changes in interconnections over time it is however more promising to look at dependencies at the level of assets and their contributions to the comovement of markets and sectors. Networks can describe such large  systems and they can also serve as a dimensional reduction.

 % This paper proposes a mapping of interconnectedness in the global stock market. 
 Interconnections between financial markets play a dual role \citep[see also][]{martinez2019interconnectedness}. On the one hand, they can absorb shocks and lead to greater robustness. They can also propagate shocks and create greater fragility. As such, measuring interconnectedness is of value and interest. The methods however, often depend on the availability of data and the kind of underlying financial or economic activity. Thus, measuring interconnections ranges from the use of direct exposures data, usually available only to supervisory and regulatory entities, to publicly available data like market prices. One popular approach, which we also follow, is to investigate the empirical correlation of asset returns and the resulting implied network structure. The resulting networks help to describe the financial system as a whole, its systemic structure, and flesh out possible contagious channels and transmission mechanisms. Thus, we focus on a measure of statistical interconnectedness, based on the econometric analysis of comovement patterns \citep[see also][]{Billio2012,diebold2014network}, which is modified to address a large sample size and cross sectional variations. 
 
% The comovement of stocks can be summarized by a country- and sector-wise grouping of stocks, and this approximation provides information for strategic portfolio decisions and risk management. The paper also presents a framework to monitor the changes in interconnectedness within and between markets, and to identify transmission channels and vulnerabilities. 

The results from this study contribute to the debate about the stability of interdependencies of financial markets and provide evidence of rather volatile relationships \citep[see also][]{corsetti}. Finally, even though the objective of the proposed methodology is not a classical factor model, we also discuss the presence of sectoral effects. We observe that these effects are difficult to model and describe only a limited amount of overall connectivity across stocks, which supports the findings for example presented by \cite{forbchinn}.  

Apart from analyzing stock comovement, this paper builds on two additional strands of literature. First, it is related to the macroeconomic literature that discusses the effects of market openness on business cycle synchronization \citep[see, e.g.,][]{brooks,opencycle,claessens2018frontiers}. These effects depend on similarities in industry structure, although this is often overshadowed by country-specific effects \citep{imbs}.  There are also studies on the transmission of shocks in a crisis situation. In such a situation, the determinants for spillovers can change relative to what is observed in normal times. See, for example, \cite{fratzscher} for an analysis of the 2008 crisis and \cite{kaminsky} for an analysis of crises in the 1990s. 

Second, this paper is also related to the literature of network-based analysis of financial markets. In the economics literature the analysis of networks has gained importance since the 2008 financial crisis and the analysis of cascading effects in the financial sector. \cite[see][]{summer2013financial,gaiandandy,glasserman2016contagion}. The comovement of stocks as a network has been analyzed for some time by interdisciplinary researchers. In these approaches, the similarity of stock returns is interpreted as information about linkages between stocks and networks are produced either by applying planar filtering or correlation thresholds \citep{structure,Gopikrishnan,plos}. These approaches deal well with the complexity of financial markets and have analyzed cases where the dimensionality is high. A weakness of these studies is that they are often of a more exploratory nature and that the number of links is often chosen ad hoc. We contribute to this literature by proposing an approach where the dimensionality is also high but where each of the connections in this network can be quantified by strength and significance, thus connecting the literature on networks to the literature in financial econometrics.

%We investigate a dataset of nearly 4,000 stocks from 15 countries. We analyze the dependencies of these stocks by assessing significant dependencies on the individual stock level. These dependencies are aggregated hierarchically by sector and country. The properties of network representations of these dependencies are then analyzed for structural properties and their dynamics. The methodological framework presented here can quantify the evolution of interconnectedness in the global market over time, and shed light on the importance of sectoral factors and the extent of remaining regional segregation. The paper also shows aspects of market interconnections can be overlooked by using market indices only. We show that while the overall level of dependence is volatile, the relative importance of specific countries and sectors shows some stability. We test different hypotheses for segmentation in the global market and find that regional clustering still offers most explanation for the network structure. We also find sectoral effects, mainly stemming from the Energy, Materials and Financial sector. However, these affects are only visible in the microscopic structure of the network.

The remainder of the paper is organized as follows: Sections \ref{sec:data} and \ref{sec:methods} describe the stock market data set, the methodology to measure interconnectedness in terms of dependencies between stocks, and how this information can be used to derive networks. Static and dynamic networks and an analysis of their properties are presented in Sections \ref{sec:static} and \ref{sec:dynamic}, respectively. In Section \ref{sec:ergm} we study the structure of the derived networks, and investigate the presence of geographic or sectoral factors.

\section{Data}\label{sec:data}

The data used in this study consist of the daily closing prices of stocks listed on exchanges in Australia, Brazil, China, Spain, France, the United Kingdom, Hong Kong, India, Japan, South Korea, the Netherlands, Singapore, United States, Canada, and Germany. We chose stocks that were components of a benchmark index, and were continuously traded with sufficient volume throughout the sample period.\footnote{Excluded were stocks that were exempt from trading for more than 10 days in a row as well as stocks without price movement and/or negligible trading  volume for more than 8 percent of the total trading days. Also excluded were 5 stocks which at some point showed price behavior or market capitalization similar to penny stocks. Twin stocks  were removed.} Some countries had a large number of stocks that met our criteria. For this reason, we selected the 500 stocks with the highest market capitalization from exchanges in India, Japan, China, and Korea. We chose U.S. stocks that are in the S\&P 500 index. For the UK, we selected stocks in the FTSE 350.\footnote{Data were obtained from Standard \& Poor's Compustat and Thomson Reuters Corp. Prices are in local currency. 'China' refers to mainland China and stocks that are traded within mainland China. For the SAR of Hong Kong stocks are comprised of local stocks as well as other Chinese stocks which legally reside in Hong Kong. A list with the names of all stocks is available upon request.} The Global Industry Classification Standard (GICS) sector designation for each stock was used, when available. When GICS was not available, we used the TRBC classification from Thomson Reuters. The number of stocks by country and sector is summarized in Table \ref{tab:sectors}.

\begin{table}[h!]
\setlength{\tabcolsep}{3pt}
\small
\begin{center}
\begin{tabular}{l| c c c c c c c c c c c} 
%\hline
 Country &  Energy  & Material.  & Indus.  & Cons.D.  & Cons.S. &  Health &  Finan. &     IT  & Telec.  & Util. \\
\hline
AUS &     12 &     21 &    25 &     17 &     7 &      8  &    29 &      5 &      2  &     3 \\
BRA &      2 &     11 &    10 &     13 &     5 &      1  &     5 &      3 &      2  &    14 \\
CHN &     10 &     95 &   134 &     90 &    36 &     42  &    36 &     40 &      0  &    17 \\
ESP &      1 &      9 &    15 &      6 &     4 &      4  &    11 &      3 &      1  &     6 \\
FRA &      6 &     15 &    47 &     53 &    16 &     18  &    42 &     47 &      4  &     9 \\
GBR &     17 &     19 &    69 &     54 &    21 &     10  &    68 &     26 &      3  &     8 \\
HKG &      1 &      2 &    16 &      8 &     2 &      1  &    37 &      3 &      2  &     6 \\
IND &      7 &    125 &   105 &    116 &    37 &     28  &    21 &     51 &      4  &     6 \\
JPN &      3 &     35 &   125 &    115 &    20 &     17  &    24 &    157 &      3  &     1 \\
KOR &      2 &     91 &   101 &    102 &    27 &     35  &    11 &    125 &      4  &     2 \\
NLD &      2 &      3 &    17 &      8 &     9 &      0  &    12 &      9 &      1  &     0 \\
SGP &      0 &      0 &    11 &      3 &     1 &      1  &    18 &      2 &      2  &     1 \\
USA &     38 &     26 &    65 &     73 &    43 &     43  &    81 &     55 &      6  &    31 \\
CAN &     50 &     43 &    19 &     13 &    11 &      3  &    40 &      6 &      6  &     8 \\
GER &      6 &     11 &    50 &     29 &     8 &     22  &    15 &     33 &      5  &     4 \\
\hline
\end{tabular}
\caption{Number of stocks by sector and country. Sectors are: Energy, Materials, Industrials (Indus.), Consumer Discretionary (Cons. D.), Consumer Staples (Cons. S.), Health Care (Health), Financials, Information Technology (IT), Telecommunication Services (Telec.), and Utilities (Util.). Countries are: Austria (AUS), Brazil (BRA), China (CHN), Spain (ESP), France (FRA), United Kingdom (GBR), Hong Kong (HKG), India (IND), Japan (JPN), South Korea (KOR), Holland (NLD), Singapore (SGP), United States (USA), Canada (CAN), and Germany (GER).}\label{tab:sectors}
\end{center}
\end{table}

Our sample period starts on 1 July 2006 and ends on 30 June 2013. The eight years of data results in $T=1329$ trading days and $N=3828$ stocks. For stocks in all countries, we use data from trading days when the stock markets in London and New York are both open. Because the stocks are traded in different time zones there are limitations in the synchronization of the returns. The international date line in the Pacific necessitates that Asian countries finish trading first in the day, while the Americas finish last. The correlation of daily returns will naturally under-represent the amount of comovement between the most distant countries because trading takes place without an overlap in time. To address this issue, we use two approaches: To analyze the long-run effects, we calculated weekly returns for all time series, which leaves us with $T'=365$ observations. To analyze short-run effects, we calculate a correction factor to use with the daily data; details are explained in Section \ref{sec:time}.

The remainder of this paper uses the returns time series derived from the log price changes of the stocks, $r_t = log(p_t) - log(p_{t-1})$. The number of stocks per country varies from 39 for Singapore to 500 for the larger countries (see summary statistics presented in Table \ref{tab:stats}). The markets in China, Korea, and India have imposed limits on the maximum daily price changes. In Korea, for example, the limit for the daily price movement was 15 percent. The distributions of the returns time series for these countries are truncated. This does not mean that the volatility is necessarily lower (see sample variances in the table). All time series of asset returns are heavy-tailed, as the values for the kurtosis indicates. The tail exponent was calculated with the Hill estimator, and the values are mostly slightly greater than 3.

To uncover dependencies between stocks in different countries by sector, it is necessary to test if stocks within a specific sector are more correlated than stocks from random sectors. The far right columns in Table \ref{tab:stats} show the results from calculating the average of all within-sector correlations, the average of all between-sector correlations, and the ratio between the two. The average of the first is significantly higher for all countries except for China. The dispersion and level vary by country. In Japan, only two sectors show a higher than average between-sector correlation than the within-correlation (see also Figure \ref{fig:sec_corr} in the appendix). For the case of China this behavior can mostly   be explained with the large influence of the state and significant limitations for foreign investors. Japan, on the other hand is special case among the developed markets. The presence of large horizontally integrated conglomerates and a slightly more inward looking economy are likely the reasons for its unique market structure \citep[see also][]{china,lincbook}.

\begin{table}
\setlength{\tabcolsep}{3pt}
\begin{center}
\small
\begin{tabular}{c| c  c c c c c c }

    &   $N$          & $var(r)$  & Kurtosis & Tail       & Avg. correlation & Avg. correlation & Correlation\\
		&   stocks             &           &           & exponent  & within sector & all stocks & Ratio\\
		Country\\
		\hline
AUS &  129 &       0.00064 &   18.7 &  3.47 &  0.35 &  0.18 & 1.94 \\
BRA &   66 &       0.00063 &   14.4 &  3.77 &  0.42 &  0.22 & 1.91\\
CHN &  500 &  0.00105 &    5.0 &   --     &  0.27 &  0.25 & 1.08\\
ESP &   60 &       0.00057 &   12.1 &  3.61 &  0.38 &  0.23 & 1.65\\
FRA &  257 &       0.00059 &   26.4 &  3.32 &  0.22 &  0.14 & 1.57\\
GBR &  295 &       0.00064 &   25.6 &  3.40 &  0.22 &  0.16 & 1.37\\
HKG &   78 &       0.00082 &   18.0 &  3.35 &  0.44 &  0.20 & 2.20\\
IND &  500 &  0.00114 &    8.5 &   --     &  0.31 &  0.25 & 1.24\\
JPN &  500 &       0.00120 &   15.1 &  2.99 &  0.37 &  0.31 & 1.19\\
KOR &  500 &  0.00124 &    8.0 &    --    &  0.36 &  0.27 & 1.33\\
NLD &   61 &       0.00059 &   25.7 &  3.38 &  0.40 &  0.25 & 1.60\\
SGP &   39 &       0.00052 &   38.1 &  3.37 &  0.48 &  0.29 & 1.65\\
USA &  461 &       0.00065 &   24.2 &  3.25 &  0.24 &  0.21 & 1.14\\
CAN &  199 &       0.00070 &   21.1 &  3.26 &  0.23 &  0.15 & 1.53\\
GER &  183 &       0.00074 &   17.1 &  3.44 &  0.25 &  0.15 & 1.66\\
\hline
\end{tabular}
\end{center}
\caption{Statistics for the Returns Time Series. We calculated the variance, kurtosis and the tail exponent from all returns in each country. In three of the markets, the maximum daily price change is constrained (caps), which permits the analysis of the tail exponent. The far right columns show the average correlation of stocks within a given sector is always larger than the average correlation of all stocks within a given country.}\label{tab:stats}
\end{table}

\section{Estimating inter-market and intra-market interconnectedness}\label{sec:methods}

For the following, it is important to note that the analysis of comovement across markets differs from the well understood properties of comovement within markets \citep[see, e.g.,][]{barberis,green}. In the latter, one describes the behavior of individual stocks. This is not the case in the analysis of comovement across markets. Most approaches to studying comovement on the global level have focused on the analysis of stock market indices \citep[see, e.g.,][]{baur} or other smaller samples including sectoral indices. A wide range of methods has been applied, among these are unit root and cointegration tests, vector autoregression models, correlation-based tests \citep{contagion,fry}, causality tests \citep{Billio2012}, multivariate GARCH models \citep{ddcg}, and models of variance decomposition \citep{yilmazvar}. 
While advances have been made in applying these models to larger data sets they are still impractical for cases like ours \citep[see also][]{aielli,large_dcc2,largevar}. 

\subsection{GARCH filtering process}

Our approach is based on pairwise comparisons of stock returns.
Although an analysis of the correlation of these returns can be informative, changes in volatility in all time series would govern the results, since volatility in financial markets around the world is largely synchronized \citep[see][]{meteor}. The long memory in volatility would also complicate the assessment of significance bounds \citep[see also][]{contagion}.

We choose to investigate dependencies between stocks on a more general level and therefore apply a filtering to the returns. As a filter we use the conditional variance of a univariate GARCH model \citep{Boll}. This process removes volatility changes and autocorrelation from the returns. 

This means that we assume that the returns follow a random process with \mbox{$\varepsilon_t = v_t \sqrt{h_t}$},
where $v_t$ is white noise and
\begin{eqnarray}
 h_t = \alpha_0 + \sum_{i=1}^q \alpha_i \varepsilon_{t-i}^2 + \sum_{i=1}^p \beta_i h_{t-i}.
\end{eqnarray}
We will make use of the conditional variance $h_t$ to calculate filtered returns\footnote{It should be noted that also the covariances of $h_t$ can be used to analyze interconnections, a short comparison of these two measures can be found in \ref{app:corr}.} such that 
\begin{eqnarray}
r_{i,t}^{f} = \frac{r_{i,t}} {\sqrt{h_{i,t}} },
\end{eqnarray}
for all stocks $i$. We obtain time series with unit volatility.\footnote{Except for a few exceptions for stocks from developing markets, the GARCH(1,1) model fits very well and yields the expected coefficients for $\alpha$ and $\beta$. We checked the robustness of our results by omitting the 83 stocks for which the fit of the GARCH model was least satisfactory, but could not find any significant change on terms of the resulting p-values as described in section \ref{sec:net} \citep[see also][on robustness]{hanslund}. }

\subsection{Estimation and correction for non-synchronous trading}\label{sec:time}

By ``de-garching'' the returns, a time series is obtained that can be treated in an almost standard regression framework. Running a pairwise regression of all the filtered returns generates a measure for the comovement. The only econometric issue of these time series is that the residuals are not normally distributed, which we account for by using a robust regression \citep{robust} with t-distributed errors.\footnote{Note that the sample size would in principle allow to proceed from here with a standard regression analysis or the calculation of correlation coefficients. However, since we need reliable standard errors and significance bounds we rely on a robust regression framework. See also Figure \ref{fig:distr} in the appendix for details on the distributions of returns and residuals.}

To measure interconnectedness, we estimate pairwise the dependencies for all pairs of stocks $(i,j)$ 
	\begin{eqnarray}
	r_{i}^{f} = \beta_{0,ij} + \beta_{1,ij} r_{j}^{f} + \epsilon.
	\end{eqnarray}
Next, we focus on stock-to-stock relationships that are significant with respect to a certain threshold. Therefore, in the following we will use the p-values that can be obtained from this estimation and refer to the elements $p_{ij}$ of the matrix of p-values. Its rows and columns are ordered by countries.\footnote{For the time horizons in this study, we do not observe significantly negative correlation.} 

To test the robustness of these results, which depend on the univariate de-garching, we compare our results with those of a multivariate GARCH model. These models can only be estimated with a limited number of time series. We use the Dynamic Conditional Correlation (DCC) model \citep{ddcg} with pairs of stocks from the sample and compare the average correlation implied by the DCC model with the correlation of our filtered returns. As expected, the results are indistinguishable for long windows. For windows of 190 days, the results on the stock level differ, but the differences are marginal and unsystematic (see \ref{app:dcc}). We have also analyzed the distributions of our p-values and run simulations with permutations of our returns data to check for sensible significance levels and possible problems due to a multiple comparison problem. We found that the distributions of p-values are well-behaved and that an inflation of false positives caused by the large sample size is not to be expected, for details see figure \ref{fig:pvals} in \ref{app:dcc}. 

A challenging issue is the time difference in trading hours across markets. It leads to problems in determining the true dependencies between stocks from different markets. For stocks from Europe, this is a minor issue because UK trading time differs from other European countries by only one hour. The U.S. market opens before EU markets close, for a difference of six hours. Calculating the dependencies between stocks traded in the Americas and Asia is most problematic. The dependencies are biased downwards because the time series are asynchronous \cite[see also][]{martens}. In general, this can be dealt with in two ways: by using tick data and calculation of synchronous pseudo closing prices, or by time aggregation. The first method would require massive amounts of data and even then it would be difficult to find a specific time each day when (pseudo) price quotes for all stocks would be available. The second method is easier but limits the time resolution for our analysis. 

In the following we will use weekly returns when we analyze the entire time period. An analysis of the dynamics of dependencies (on the time scale of months) however requires the use of daily data.  We therefore propose a method that uses the difference between the average estimated daily and the weekly correlations to calculate a correction factor that ties the estimation results for the daily data to the weekly data results. This procedure is related to the works by \cite{christensen} and \cite{hayashi}, albeit much simpler and to be understood as an approximation.

We calculate a correction factor for the p-values in the following way: let $\bar{P}^w$ and $\bar{P}^d$ be $N\times N$ matrices where the elements are the average p-values on a country-to-country level (the matrix contains blocks with identical values resembling all pairs of countries). After adding 1 to each element of these matrices, we can calculate the element-wise (notation:$\: ./ \:$) ratio of the p-values of the weekly and daily estimates.
\begin{eqnarray}
R =  (1+\bar{P}^w) \: ./  \: ({1+\bar{P}^d}).
\end{eqnarray}\label{eq:pcor}
Then, a matrix with correction factors $C$ can be calculated,
\begin{eqnarray}
C = min \: \bigl( \: 1 \:, \: R + 1 - \langle diag^*(R) \rangle  \bigr),
\end{eqnarray}
where the last term corrects for differences in the p-values that are not due to non-synchronous trading times (we denote by $diag^*$ the diagonal blocks of $R$ capturing the dependencies between stocks within one country).  The corrected p-values can then be obtained as
\begin{eqnarray}
P = max \: \left( \: 0 \:, \: \left( (1+P^d) \circ C \right) -1 \right) .
\end{eqnarray}
The $min$  and $max$ ensure that all p-values stay within the $[0,1]$ band.

\begin{figure}
\begin{center}
	\includegraphics[width=0.7\textwidth]{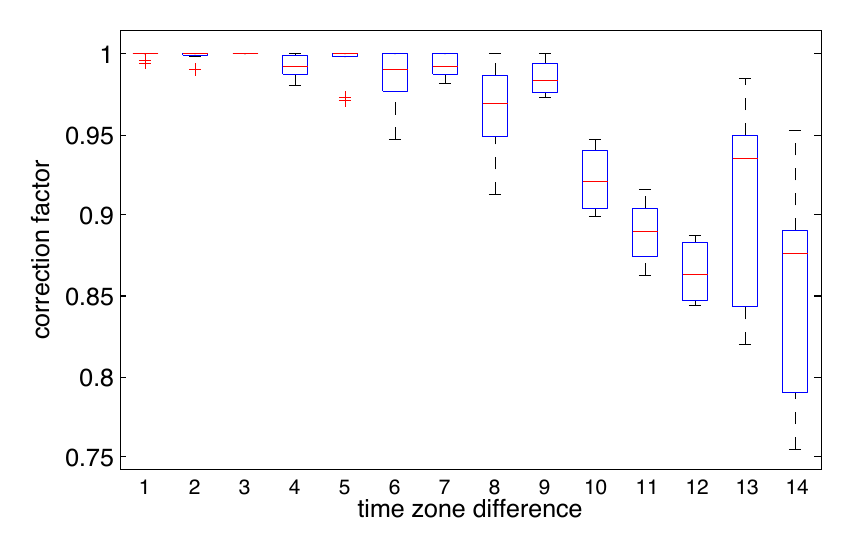}
   \caption{\label{fig:time}Range of Correction Factors by Time Zone Differences. The plot shows the difference between the average estimated dependency of stocks on the daily level versus the weekly level (see eq. \ref{eq:pcor}). Significant differences between the estimated dependencies arise when the markets are more than three time zones apart. Less-developed markets with below average comovement show a lower  difference than developed markets (this explain most of the dispersion for the two far right box-plots).}
\end{center}
\end{figure}

To summarize, this procedure assumes that overall estimates from the daily and the weekly data are similar and remaining differences are likely to be a result of non-synchronous data. The ratio of the slightly transformed p-values is used to correct for this issue so interdependencies that were otherwise discarded as, say just below the 90 percent confidence bound, will now be accounted for as significant if the correction factor for this country pair is sufficiently smaller than 1. 

The calculated correction factors are presented in Figure \ref{fig:time}. As expected, the correction increases with geographical distance, even though this does not fully explain the issue. The correction remains relatively small for distant, less-developed markets where the comovement is low anyway (which explains the wide range of the far right boxplots).

\subsection{Visualizing and quantifying interconnectedness}\label{sec:net}

\begin{sidewaysfigure}
\begin{center}
\includegraphics[width=0.99\textwidth]{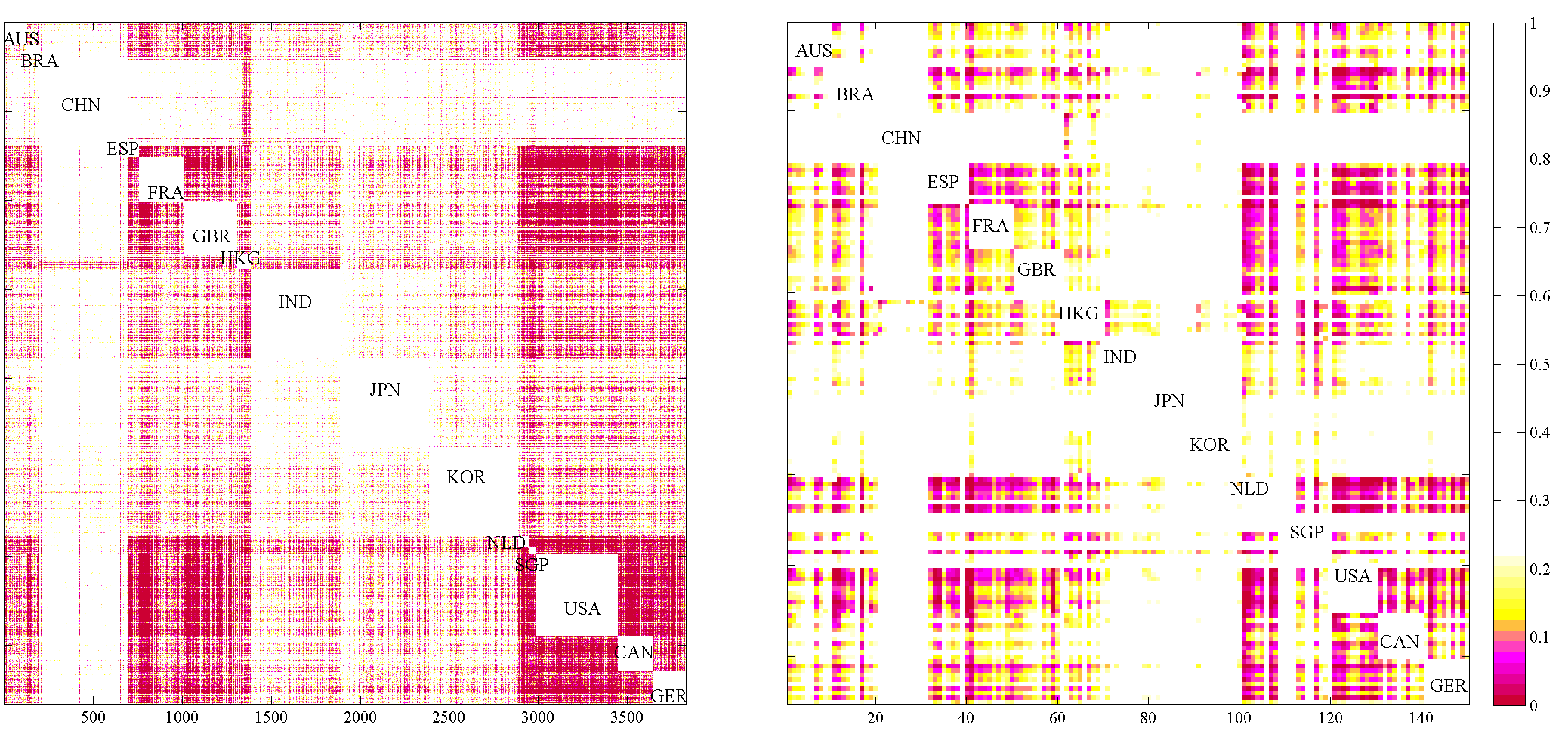}
   \caption{\label{fig:cmap} p-Values for the Stock-to-Stock Dependencies (left) and Sector Averages (right). The stocks are arranged by countries and the abbreviated country names are plotted along the main diagonal. Entries for dependencies within countries have been removed. The values are color-coded with dependencies above the 90 percent confidence interval appearing pink to red. In the right panel, the values are averaged on a sector to sector basis, so that for each country we show the average dependency to the 10 sectors in the 14 other countries. This reduces the matrix to 150 $\times$ 150 entries.}
	\end{center}
\end{sidewaysfigure}

In a financial network, entities such as banks, financial institutions, central counterparties, and traders are considered nodes. Their relationships in interbank lending, contractual obligations, and counterparty exposures define the links that connect them. Stocks and their relationships can also be expressed as a network. Each stock is a node and its estimated interdependencies show whether, and how strongly, the nodes are connected to each other. This information is stored by using adjacency matrices, where the entries in row $i$ and column $j$ indicate the strength of the connection between the respective nodes (see Figure \ref{fig:cmap}).

Because the p-values contain this required information and their distributional properties allow a useful weighting, it is relatively easy to obtain adjacency matrices.
The matrices of p-values are converted to adjacency matrices $A$ by first removing all entries where the significance level for the stock-to-stock dependence is below a certain threshold $\gamma$. In the following, we use $\gamma < 0.1$ if not stated otherwise.\footnote{Our qualitative results do not depend on the particular threshold chosen, it is however necessary to allow for a certain range of p-values such that within-market as well as in-between-market interdependencies enter the following analysis.} The estimated significance level can be used as a measure of connection strength by defining
\begin{eqnarray} 
A_{ij} \propto (\gamma - p_{ij}) .
\end{eqnarray}
This adjacency matrix has a weighted positive entry if stocks $i$ and $j$ are significantly linked, measured by the estimated conditional correlation. This procedure of creating the adjacency matrix is in fact similar to the method proposed by \cite{Billio2012}. Their p-values are however the outcome of a linear regression of raw returns in a Granger-causality analysis.\footnote{Note that Granger-causality should not detect any significant relationships in efficient financial markets. Hence, while its use in the above mentioned paper can be justified by market frictions in the time of crisis, it would be less useful in our analysis of comovement, since it would be impractical to dis-entangle the influences that lead to positive results, let alone to compare them over time.}

%To uncover the dependencies between the markets on a sector level, the stock level dependencies can be used to describe the resulting network on a sector to sector basis. 

The averages of the p-values of the relationships of the stocks in a specific sector in one country are then used with stocks in a specific sector in another country to map  the sector-by sector dependencies.\footnote{The averages of the p-values are  not p-values anymore. However, in this case it makes sense to use these averages because of low dispersion within the groups of stocks. Also note that in Table \ref{tab:indexcorr} median values are used for the comparison because the distribution of p-values for all stocks within one country is slightly skewed. On the sector level, however, the difference between the mean and the median is small, the latter value is mostly slightly higher and would tendencially lead to more significant links.} The translation into a corresponding adjacency matrix is done in the same way as before but the dimension is reduced to $150\times150$ (15 countries, 10 sectors). Sectors in countries that consist of fewer than two stocks are excluded from further analysis.

Finding useful visual representations of adjacency matrices is a complex process and the equivalent of finding a good dimensional reduction of a N-dimensional system, where N is the number of nodes. This process is also related to the problem of community detection in graphs, which is a high-dimensional clustering problem.\footnote{See also the book by \cite{newman} for an in-depth introduction to Network Science.} We graph our networks by applying the widely used algorithm developed by \cite{hu}. This algorithm arranges nodes in a two-dimensional space in such a way that the total edge length of the graph is minimized, which shows the most pronounced communities of nodes within a graph. This specific algorithm uses the physics of repulsion to generate a visualization.\footnote{We have verified that the qualitative results of the obtained visualizations do not depend on the choice of this specific algorithm.}

In order to investigate the determinants that govern the connectivity in these networks we estimate different exponential random graph models \citep[ERGMs, see][]{strauss}. In these models the links and non-links between stocks are the dependent variable and we estimate the influence of fixed effects on the country and sector-level as well as interaction effects on the likelihood of a link between pairs of stocks. While the estimation results can be interpreted similarly as those of a logistic regression, for numerical reasons these models employ simulated maximum likelihood.\footnote{For an in depth treatment of ERGMs we refer the reader to \cite{ergmbook}.}

\section{Static analysis of global interconnectedness}\label{sec:static}

\begin{figure}[p]
\begin{center}
\includegraphics[width=\textwidth, trim = 105 10 90 10, clip=true]{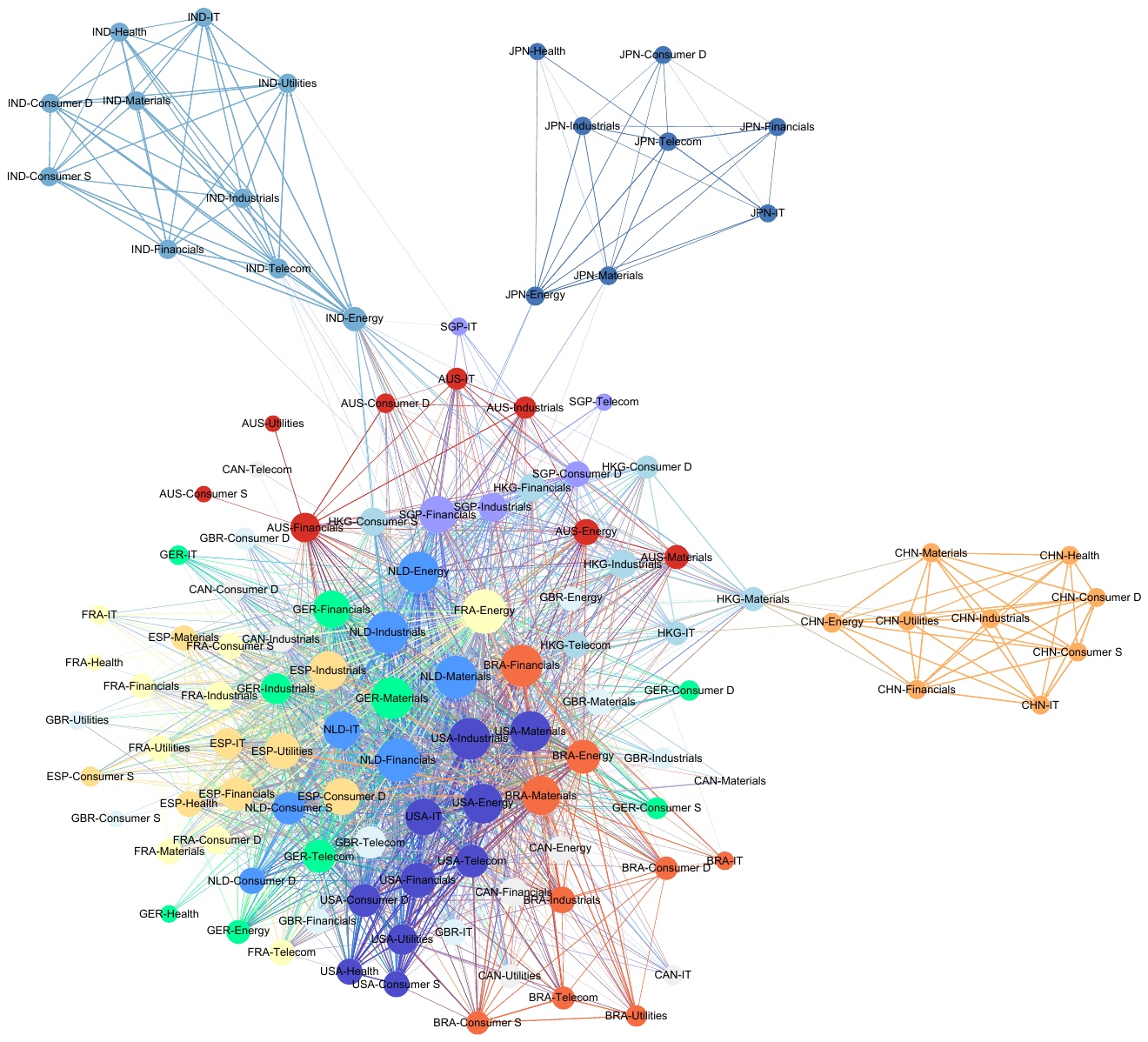}
   \caption{\label{fig:weekly_sec} Network Representation Based on Sector-wise Averaged Estimated Dependencies from Weekly Data.
	Nodes (sectors) from the same country have the same color. The node size is proportional to the degree of the number of significant links/dependencies. The sector network displays features similar to the stock network in Figure \ref{fig:cmap}. Most sectors form a central cluster around the U.S. market. India, China, and Japan form loosely connected cliques. Sectors with few or no links are omitted (South Korea is not present). The layout was performed in Gephi using the Yifan Hu algorithm.}
	\end{center}
\end{figure}

We first apply the proposed methodology on the entire sample period, to which we refer as a static analysis. The results obtained from analyzing the weekly returns are depicted in Figure \ref{fig:weekly_sec}. Stocks from Western markets form a hairball in the middle of the network, showing they are highly interconnected. Within this hairball, the mixing is strongest within the European markets, while otherwise regional structures remain visible. The markets of China, Japan, and India are only loosely connected to the central component of the network.
These results are in line with earlier studies based on market indices by, for example, \cite{det_inte} and also with earlier network-based approaches like \cite{kenett2012correlations}, where simpler correlation measures were used. The new approach however now allows to observe the interconnectedness across especially the Western countries in more detail. In particular, the proposed methodology provides the means to identify cross-market relationships on a finer grain, namely to identify specific sectors and their centrality in the financial network. We have compared these results based on sectors with those for the stock level interactions (see Figure \ref{fig:weekly} in the appendix), and have verify that the aggregation procedure does not influence our findings.

It is possible to observe that even the most central sectors in the densely connected middle of the network are still mostly grouped next to sectors from the same market, which is a sign of remaining regional segmentation. When we search for exceptions from this pattern we find examples like the German materials sector or the British and French energy sectors. These nodes are surrounded by related sectors from other markets which hints that in some instances sectoral effects can be as strong as market or regional effects.

\begin{figure}[htb]
\begin{center}
\includegraphics[width= \textwidth, trim = 20 300 10 310, clip=true]{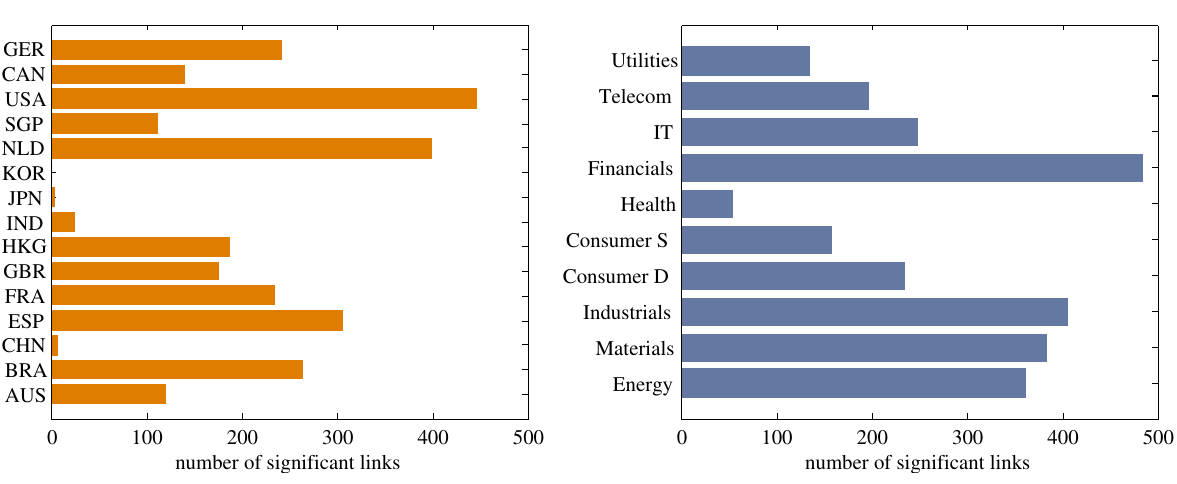}
   \caption{\label{fig:nlinks} Number of Significant Links by Country (left) and by Sector   (right). We count the number of significant links between sectors and aggregate by country and by sector based on the estimation of the weekly data. The United States and European countries (which are slightly favored by their large number) are the most connected ones. The four Asian countries (excluding Singapore) are least connected. When aggregated by sectors, stocks from the financial sector, followed by the materials sector and energy sector, are the most connected stocks.}
	\end{center}
\end{figure}

The connectivity in this network can be summarized by counting the number of links between sectors and between countries. The results are shown in figure \ref{fig:nlinks}. The left panel confirms the visual impression of a very connected U.S. stock market. Most sectors in European markets comove in a significant manner. The right panel shows clear differences on the sector level. Stocks from the financial, industrials, materials, and energy sectors show more interconnections than stocks from other sectors, and this is consistent for all countries.

While we obtain qualitatively similar results from the stock versus the sector level, we can show that this not true if one reverts to stock market indices.
 We illustrate this by applying the same estimation methodology to stock market indices for the 15 countries. The weekly index returns are de-garched and  p-values of the pair-wise estimated dependencies are presented in Table \ref{tab:indexcorr}. The results are compared with the median of the p-values for the stock-wise analysis and the median when we hierarchically average within sectors and then overall sectors within a country. While the indices are all highly dependent, the stock level shows lower median p-values, indicating pronounced heterogeneity in stock-to-stock dependencies, which are in no way captured by a market index analysis. This means that the findings about market comovement in studies of indices like \cite{hamao1990correlations} should be interpreted carefully, since they refer to one particular definition of the `market'.

\begin{sidewaystable}[p]
\begin{center}
\scriptsize
\begin{tabular}{c c |c c c c c c c c c c c c c c c }

   &  &  AUS &      BRA &      CHN  &     ESP   &    FRA  &     GBR    &   HKG     &  IND    &   JPN  & KOR & NLD  & SGP  & USA  & CAN  & GER \\ 
	Country & & &\\
	\hline
& index  & 0.0000 & 0.0000 & 0.0036 & 0.0000 & 0.0000 & 0.0000 & 0.0000 & 0.0000 & 0.0000 & 0.0000 & 0.0000 & 0.0000 & 0.0000 & 0.0000 & 0.0000 \\ 
AUS & sector & 0.0823 & 0.2202 & 0.6162 & 0.1357 & 0.1220 & 0.1869 & 0.1285 & 0.3215 & 0.3196 & 0.3571 & -- & -- & 0.1190 & 0.2547 & 0.1285 \\ 
& stock  & 0.0303 & 0.1856 & 0.5673 & 0.0858 & 0.0993 & 0.1253 & 0.0718 & 0.3125 & 0.2848 & 0.3297 & 0.0481 & 0.0560 & 0.0861 & 0.1615 & 0.1058 \\ 
\hline
& index  & 0.0000 & 0.0000 & 0.0008 & 0.0000 & 0.0000 & 0.0000 & 0.0000 & 0.0000 & 0.0000 & 0.0000 & 0.0000 & 0.0000 & 0.0000 & 0.0000 & 0.0000 \\ 
BRA & sector & 0.2038 & 0.0095 & 0.6370 & 0.0530 & 0.0852 & 0.1048 & 0.0748 & 0.2759 & 0.3910 & 0.2513 & 0.0616 & -- & 0.0323 & 0.1516 & 0.0746 \\ 
& stock  & 0.1774 & 0.0073 & 0.6129 & 0.0524 & 0.0903 & 0.0977 & 0.0635 & 0.2890 & 0.3657 & 0.2749 & 0.0342 & 0.0595 & 0.0284 & 0.1043 & 0.0812 \\
\hline 
& index  & 0.0018 & 0.0001 & 0.0000 & 0.0390 & 0.0070 & 0.0040 & 0.0000 & 0.0020 & 0.0097 & 0.0006 & 0.0111 & 0.0002 & 0.0118 & 0.0021 & 0.0026 \\ 
CHN &sector & 0.5655 & 0.5435 & 0.0000 & 0.7299 & 0.6068 & 0.6202 & 0.1665 & 0.5258 & 0.6351 & 0.4136 & 0.5710 & -- & 0.6674 & 0.7197 & 0.6041 \\
 &stock  & 0.5333 & 0.5367 & 0.0000 & 0.7302 & 0.6150 & 0.6415 & 0.1964 & 0.5441 & 0.6586 & 0.4689 & 0.6305 & 0.3510 & 0.6723 & 0.6686 & 0.6219 \\ 
\hline 
& index  & 0.0000 & 0.0000 & 0.0826 & 0.0000 & 0.0000 & 0.0000 & 0.0000 & 0.0000 & 0.0000 & 0.0000 & 0.0000 & 0.0000 & 0.0000 & 0.0000 & 0.0000 \\ 
ESP & sector & 0.1344 & 0.0570 & 0.7534 & 0.0011 & 0.0163 & 0.0258 & 0.1169 & 0.2566 & 0.3311 & 0.2989 & 0.0097 & -- & 0.0044 & 0.0968 & 0.0323 \\ 
& stock  & 0.0870 & 0.0531 & 0.7557 & 0.0006 & 0.0205 & 0.0243 & 0.0825 & 0.2710 & 0.3163 & 0.2861 & 0.0031 & 0.0554 & 0.0050 & 0.0650 & 0.0275 \\ 
\hline
& index  & 0.0000 & 0.0000 & 0.0094 & 0.0000 & 0.0000 & 0.0000 & 0.0000 & 0.0000 & 0.0000 & 0.0000 & 0.0000 & 0.0000 & 0.0000 & 0.0000 & 0.0000 \\ 
FRA &sector & 0.1505 & 0.1067 & 0.6852 & 0.0236 & 0.0361 & 0.0697 & 0.1252 & 0.2882 & 0.3002 & 0.3109 & 0.0186 & -- & 0.0221 & 0.1218 & 0.0552 \\
& stock  & 0.1173 & 0.1123 & 0.6817 & 0.0287 & 0.0367 & 0.0678 & 0.0983 & 0.3047 & 0.2819 & 0.3147 & 0.0130 & 0.0712 & 0.0335 & 0.1110 & 0.0546 \\ 
\hline 
& index  & 0.0000 & 0.0000 & 0.0101 & 0.0000 & 0.0000 & 0.0000 & 0.0000 & 0.0000 & 0.0000 & 0.0000 & 0.0000 & 0.0000 & 0.0000 & 0.0000 & 0.0000 \\ 
GBR & sector & 0.2002 & 0.1315 & 0.6983 & 0.0345 & 0.0542 & 0.0526 & 0.1603 & 0.3707 & 0.3379 & 0.3612 & 0.0359 & -- & 0.0249 & 0.1469 & 0.0658 \\ 
& stock  & 0.1310 & 0.1169 & 0.6883 & 0.0304 & 0.0611 & 0.0356 & 0.1199 & 0.3596 & 0.3342 & 0.3490 & 0.0120 & 0.0876 & 0.0189 & 0.1113 & 0.0583 \\ 
\hline
& index  & 0.0000 & 0.0000 & 0.0000 & 0.0000 & 0.0000 & 0.0000 & 0.0000 & 0.0000 & 0.0000 & 0.0000 & 0.0000 & 0.0000 & 0.0000 & 0.0000 & 0.0000 \\ 
HKG & sector & 0.1316 & 0.0844 & 0.2054 & 0.0931 & 0.0936 & 0.1179 & 0.0062 & 0.1318 & 0.2378 & 0.1527 & -- & -- & 0.0729 & 0.2223 & 0.0956 \\ 
& stock  & 0.0648 & 0.0674 & 0.2215 & 0.0718 & 0.0752 & 0.1001 & 0.0007 & 0.1153 & 0.2022 & 0.1495 & 0.0348 & 0.0046 & 0.0632 & 0.1335 & 0.0741 \\ 
\hline
& index  & 0.0000 & 0.0000 & 0.0063 & 0.0000 & 0.0000 & 0.0000 & 0.0000 & 0.0000 & 0.0000 & 0.0000 & 0.0000 & 0.0000 & 0.0000 & 0.0000 & 0.0000 \\ 
IND  & sector & 0.2853 & 0.2375 & 0.6172 & 0.2144 & 0.2346 & 0.3141 & 0.1295 & 0.0018 & 0.5052 & 0.3788 & 0.1466 & -- & 0.2573 & 0.3515 & 0.2297 \\ 
& stock  & 0.2767 & 0.2573 & 0.6367 & 0.2370 & 0.2510 & 0.3057 & 0.1181 & 0.0025 & 0.5175 & 0.3890 & 0.1772 & 0.0986 & 0.2663 & 0.3224 & 0.2395 \\ 
\hline
& index  & 0.0000 & 0.0000 & 0.0448 & 0.0000 & 0.0000 & 0.0000 & 0.0000 & 0.0000 & 0.0000 & 0.0000 & 0.0000 & 0.0000 & 0.0000 & 0.0000 & 0.0000 \\ 
JPN & sector & 0.3659 & 0.3911 & 0.7258 & 0.3404 & 0.2903 & 0.3597 & 0.2810 & 0.5842 & 0.0431 & 0.4657 & 0.1923 & -- & 0.3135 & 0.4064 & 0.3113 \\ 
& stock  & 0.3023 & 0.3790 & 0.7411 & 0.3186 & 0.2822 & 0.3542 & 0.2487 & 0.5825 & 0.0401 & 0.4655 & 0.2137 & 0.1875 & 0.3062 & 0.3477 & 0.2933 \\ 
\hline
& index  & 0.0000 & 0.0000 & 0.0036 & 0.0000 & 0.0000 & 0.0000 & 0.0000 & 0.0000 & 0.0000 & 0.0000 & 0.0000 & 0.0000 & 0.0000 & 0.0000 & 0.0000 \\ 
KOR &  sector & 0.3535 & 0.2897 & 0.5265 & 0.2886 & 0.2962 & 0.3630 & 0.1710 & 0.4299 & 0.4096 & 0.1015 & 0.2323 & -- & 0.3335 & 0.4360 & 0.3005 \\  
& stock  & 0.3241 & 0.3063 & 0.5342 & 0.2932 & 0.2947 & 0.3392 & 0.1633 & 0.4333 & 0.4096 & 0.1093 & 0.2310 & 0.1574 & 0.3310 & 0.3962 & 0.3028 \\
\hline
& index  & 0.0000 & 0.0000 & 0.0211 & 0.0000 & 0.0000 & 0.0000 & 0.0000 & 0.0000 & 0.0000 & 0.0000 & 0.0000 & 0.0000 & 0.0000 & 0.0000 & 0.0000 \\ 
NLD & sector & -- & 0.0835 & 0.6929 & 0.0158 & 0.0205 & 0.0318 & -- & 0.2056 & 0.2055 & 0.2464 & 0.0156 & -- & 0.0137 & 0.0994 & 0.0309 \\ 
& stock  & 0.0625 & 0.0497 & 0.7091 & 0.0049 & 0.0135 & 0.0146 & 0.0531 & 0.2420 & 0.2191 & 0.2671 & 0.0009 & 0.0248 & 0.0045 & 0.0413 & 0.0157 \\ 
\hline
& index  & 0.0000 & 0.0000 & 0.0007 & 0.0000 & 0.0000 & 0.0000 & 0.0000 & 0.0000 & 0.0000 & 0.0000 & 0.0000 & 0.0000 & 0.0000 & 0.0000 & 0.0000 \\ 
SGP & sector & -- & -- & -- & -- & -- & -- & -- & -- & -- & -- & -- & -- & -- & -- & -- \\ 
& stock  & 0.0609 & 0.0703 & 0.4812 & 0.0586 & 0.0642 & 0.0842 & 0.0067 & 0.1251 & 0.1564 & 0.1603 & 0.0230 & 0.0007 & 0.0585 & 0.1251 & 0.0582 \\ 
\hline
& index  & 0.0000 & 0.0000 & 0.0297 & 0.0000 & 0.0000 & 0.0000 & 0.0000 & 0.0000 & 0.0000 & 0.0000 & 0.0000 & 0.0000 & 0.0000 & 0.0000 & 0.0000 \\ 
USA & sector & 0.1435 & 0.0329 & 0.7274 & 0.0048 & 0.0158 & 0.0207 & 0.0993 & 0.3015 & 0.3114 & 0.3079 & 0.0099 & -- & 0.0003 & 0.0559 & 0.0151 \\ 
& stock  & 0.0803 & 0.0278 & 0.7221 & 0.0044 & 0.0225 & 0.0145 & 0.0683 & 0.3079 & 0.2900 & 0.3152 & 0.0023 & 0.0482 & 0.0001 & 0.0284 & 0.0168 \\ 
\hline
& index  & 0.0000 & 0.0000 & 0.0056 & 0.0000 & 0.0000 & 0.0000 & 0.0000 & 0.0000 & 0.0000 & 0.0000 & 0.0000 & 0.0000 & 0.0000 & 0.0000 & 0.0000 \\ 
CAN & sector & 0.2674 & 0.1622 & 0.7475 & 0.1035 & 0.1137 & 0.1428 & 0.2351 & 0.4408 & 0.3898 & 0.4429 & 0.0982 & -- & 0.0653 & 0.1161 & 0.1157 \\ 
& stock  & 0.1646 & 0.1177 & 0.7112 & 0.0755 & 0.1044 & 0.1094 & 0.1468 & 0.3770 & 0.3431 & 0.3966 & 0.0389 & 0.1210 & 0.0368 & 0.0487 & 0.0893 \\ 
\hline
& index  & 0.0000 & 0.0000 & 0.0066 & 0.0000 & 0.0000 & 0.0000 & 0.0000 & 0.0000 & 0.0000 & 0.0000 & 0.0000 & 0.0000 & 0.0000 & 0.0000 & 0.0000 \\ 
GER & sector & 0.1544 & 0.0955 & 0.7109 & 0.0376 & 0.0618 & 0.0675 & 0.1276 & 0.2914 & 0.3006 & 0.3407 & 0.0301 & -- & 0.0238 & 0.1304 & 0.0416 \\ 
& stock  & 0.1291 & 0.1105 & 0.7107 & 0.0382 & 0.0564 & 0.0667 & 0.1052 & 0.3056 & 0.2916 & 0.3330 & 0.0151 & 0.0657 & 0.0266 & 0.1010 & 0.0423 \\

\hline
\end{tabular}
\caption{p-values for stock \emph{index} Correlations versus Median of p-Values from Stock Correlations. \emph{Sector} denominate values are obtained by first averaging over stocks within one sector, while \emph{stock} denotes the median p-value of the correlation of all stocks between two countries. For Singapore and the Netherlands, several of the sectors are not populated and therefore do not have sector results.}\label{tab:indexcorr}
\end{center}
\end{sidewaystable}

\section{Dynamic analysis of global interconnectedness}\label{sec:dynamic}

During the eight years covered by the data set, global economic and political events occurred that are likely to lead to significant fluctuations in comovement across markets. The 2008 financial crisis, the euro crisis, and Japan's 2011 tsunami are a few examples. A dynamic analysis is also necessary to show changes in global financial integration. The data resolution can characterize stock market dependencies in the order of months. We perform a rolling window approach that uses 190 days of data in 13 time steps with a 95-day overlap. We apply the same methodology as before now to daily data with a timing corrections for the p-values as described in Section \ref{sec:time}. Additionally we will look at the development of some measures for the network structure.

\begin{figure}[p]
  \begin{center}
    \subfigure[Jan 07 -- Dec 07]{\label{DN62007}\includegraphics[width=0.45\textwidth]{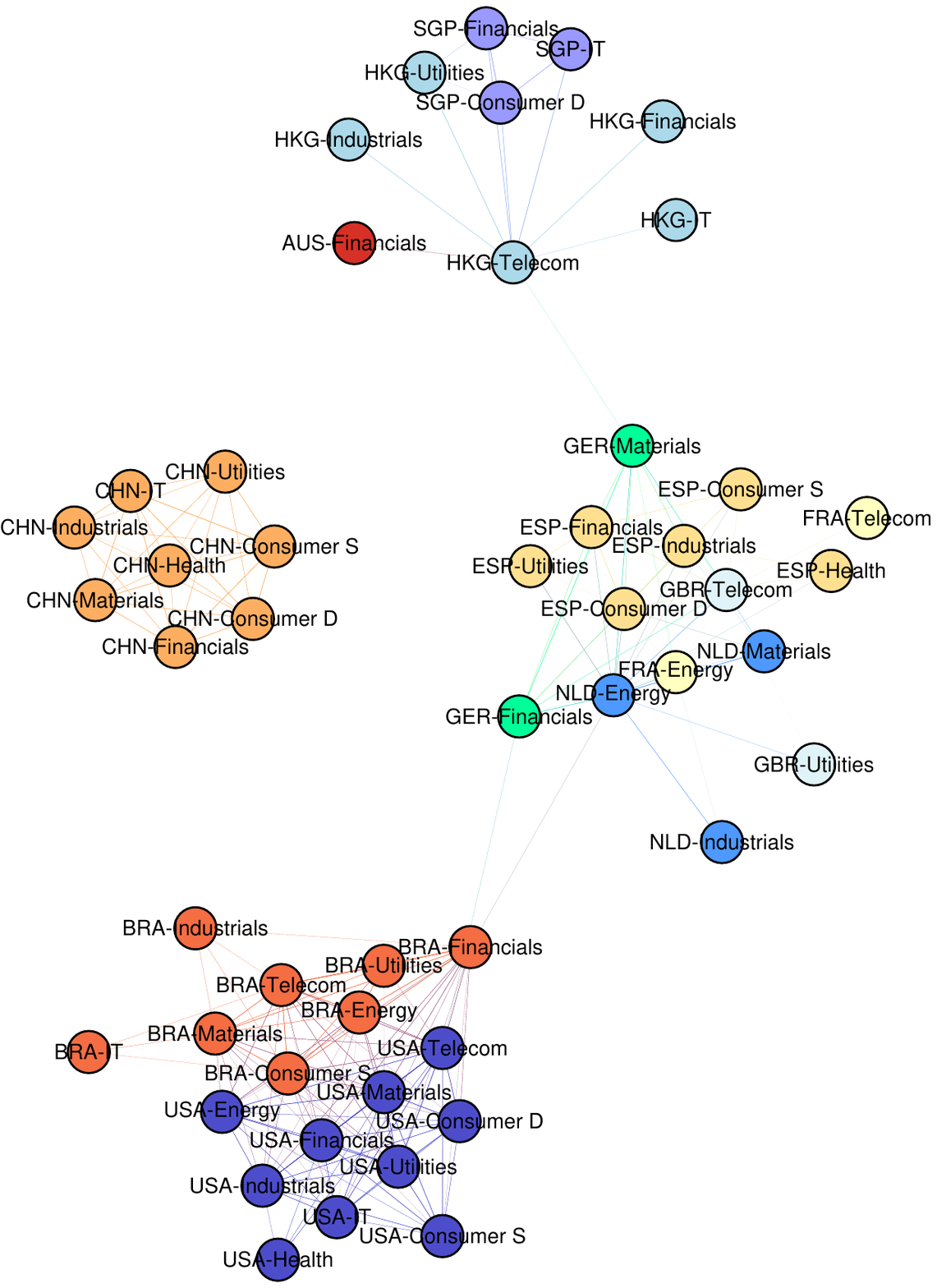}}
    \subfigure[Jul 08 -- Jun 09]{\label{DN122008}\includegraphics[width=0.45\textwidth]{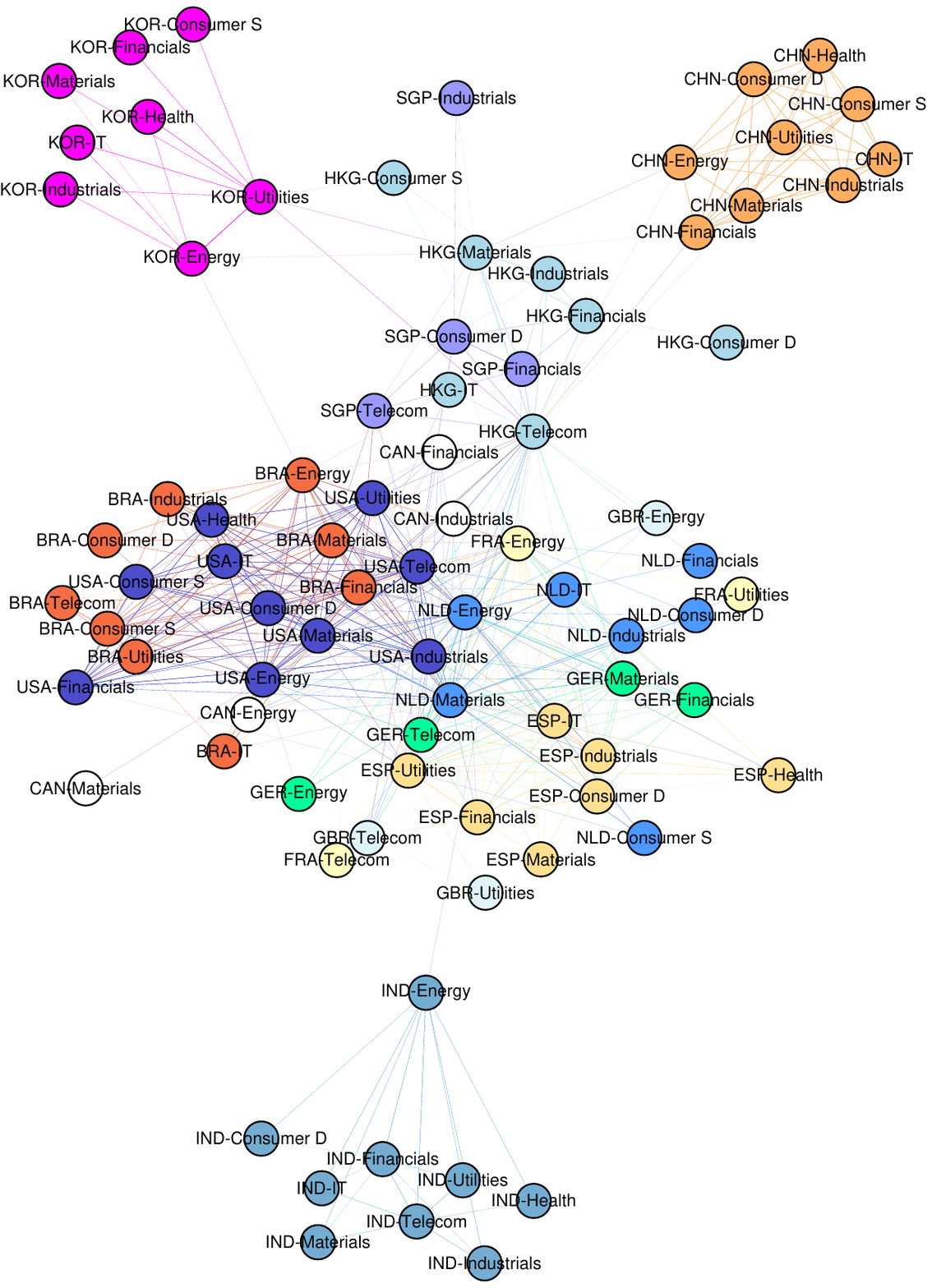}}
    \subfigure[Jan 11 -- Dec 11]{\label{DN62011}\includegraphics[width=0.4\textwidth,trim = 65 0 70 0, clip=true]{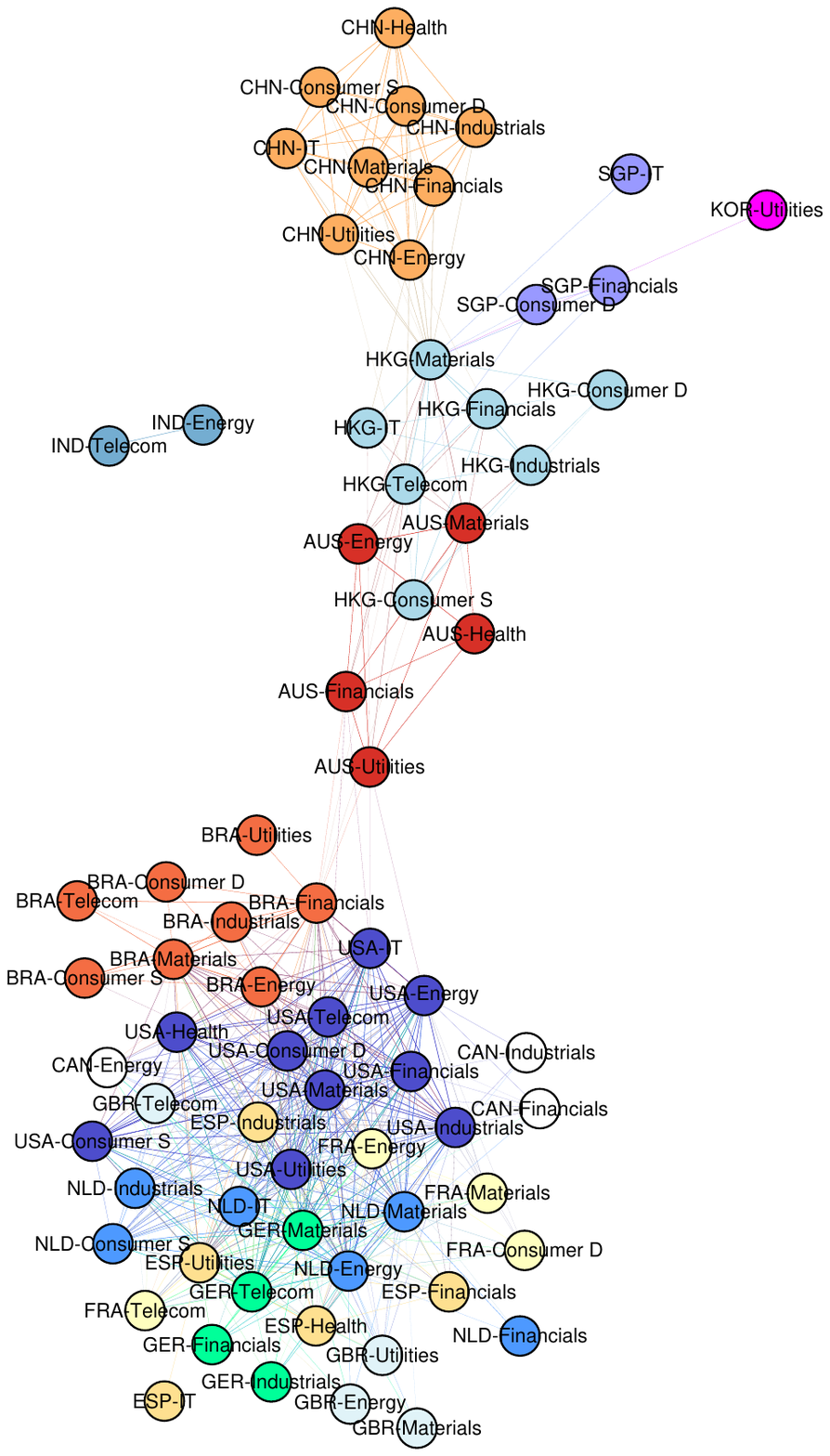}}
    \subfigure[Jul 12 -- Jun 13]{\label{DC122012}\includegraphics[width=0.4\textwidth, trim = -20 -100 -20 0, clip=true]{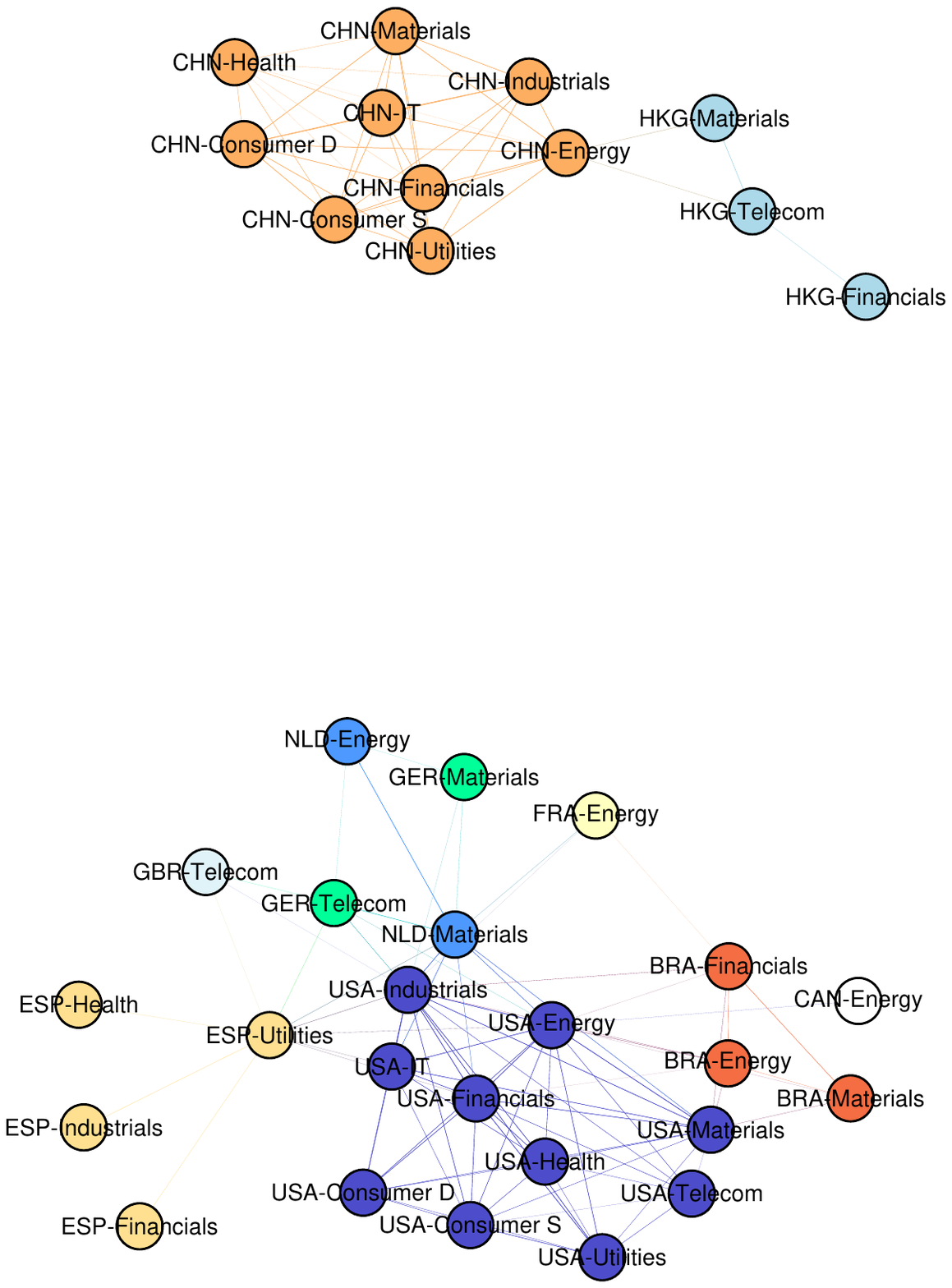}}
  \end{center}
  \caption{ Dynamics of Sector Dependency Networks. We show four networks that are representative of the dynamics in 13 time windows. Nodes (sectors) from the same country carry the same color. Links represent an average p-value of 0.1 or higher. The network shows growth and contraction during the peak of the financial crisis.}\label{fig:dmaps}
\end{figure}

Figures \ref{fig:dmaps} (a--d) display 4 of the 13 resulting networks. Interestingly these representations show pronounced deviations from the long-run representation presented in figure \ref{fig:weekly_sec}. The number of clusters, their relative position, and also the number of sectors (nodes) vary even in tranquil times. The network in the top left panel of Figure \ref{fig:dmaps} is structurally different from the others. US and European stocks form individual groups within the network, although there are multiple links between them. Some Asian markets are loosely connected to the European market. Chinese stocks show strong internal comovement but are not significantly linked to other markets.

The network changes significantly at the end of 2008 with the growing financial crisis. Stocks from all developed markets form one connected component. Some grouping into American, European, and Asian stocks remains, but the borders blur. The markets of South Korea, China, and India appear as weakly connected satellites. In fact, this network looks in many parts similar to the one presented in Figure \ref{fig:weekly_sec}, but the connections are more dense and incorporate more sectors and countries. One can argue that this time period shows strong signs of contagion, since the level of dependencies has clearly shifted upwards, compared to both the previous and the long-run state. One could further refer to this period as that of global contagion, since the stock market network is a bloomed up version of its long-run comparison.

The bottom left panel of Figure \ref{fig:dmaps} shows that in 2011 there is stronger comovement within Asia, these markets are almost separated from the European-American component. Some remaining links now run between American and Asian markets and Australia, while the comovement between European markets becomes more heterogeneous. Its nodes have also lost connections to the Asian markets. This may reflect the Euro crisis, which affected some but not all European countries. This crisis induced shifts in the structure of the network, but does not (at least at this time-scale) lead to clear signs of contagion. By 2012, the network shows comovement almost back to the pre-2008 levels. Some stock markets  are very much connected to U.S. stocks, but many sectors are no longer part of this connected core.\footnote{For comparison, networks with links aggregated on the country level are shown in Figure \ref{fig:ccmaps} in the appendix.}

\begin{figure}
\includegraphics[width=\textwidth]{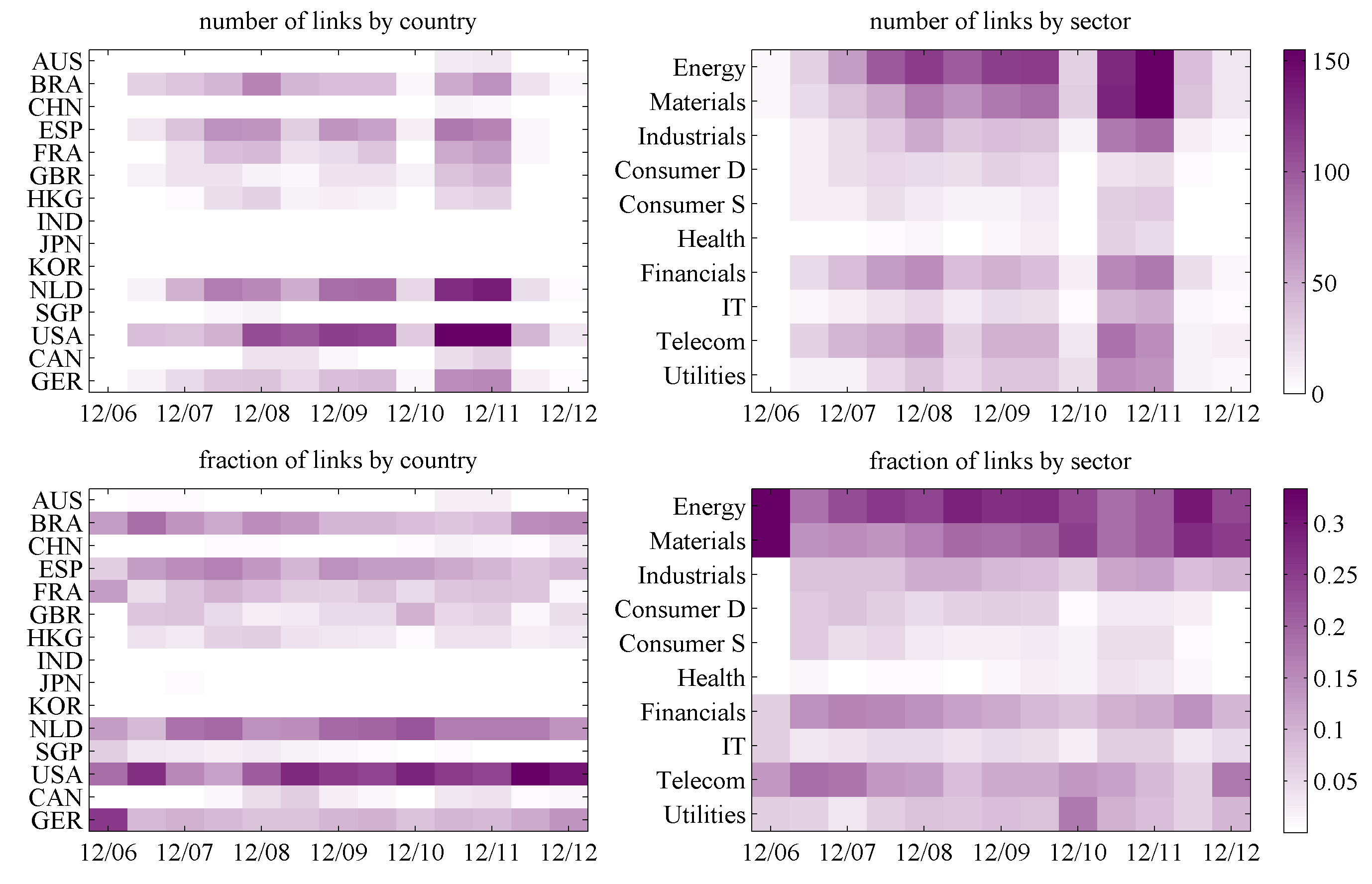}
\caption{\label{fig:dlinks} Number of Significant Links Over Time by Country (left) and by Sector (right). We count the number of significant links on the sector level and aggregate by country and by sector. The two top panels show the absolute number of links, the bottom panels show the fraction of links in the respective time window. The numbers are color coded according to the scale on the right. On the country level, we observe an overall wave-like pattern in the number of links with slight increases for the United States, Germany, and the Netherlands over time. The fractions of links by country are in fact relatively stable. Also the breakdown by sectors shows a wave-like pattern. A slight difference to the analysis of the weekly data (see Figure \ref{fig:nlinks}) is the daily data shows the energy and materials sectors as the most connected, not the financial sector.}
\end{figure}

For a deeper investigation into the origin of this variation, we study the number of links between markets on a country and sector level, similar to Figure \ref{fig:nlinks}, but with the addition of  the time dimension. The top panels of Figure \ref{fig:dlinks} show that the number of links between stock markets is low at the beginning and at the end of the sample period. It is possible to observe two peaks where many links exist between markets, at the end of 2008 and again around 2010/2011.

There appears to be some synchronization in the dynamics of the number of links, both by country and by sector. It is possible to calculate the fractions of links for each time window, which are presented in the bottom panel of Figure \ref{fig:dlinks}. This normalization can help detect shifts in the relative influences of specific countries or sectors. The country-wise view in the left bottom panel shows that the relative number of links is stable for most countries. Only the UK and Hong Kong show fluctuating behavior. The number of links for the United States is steadily increasing towards the end of the sample period. The bottom right panel with a sectoral analysis shows more interesting developments. Only in 2007 and 2008 is the financial sector a driving force for interconnections. Much more dominant are links with the energy and the materials sectors, which  gain influence throughout the sample.\footnote{Due to the low overall number of links, the values (fractions) for the first time window are noisy and discarded from our discussion.} Stocks related to consumer goods become less  important over time and the health industry is not  relevant for the entire sample period.

The ranking of the most-connected sectors varies with  the data frequency. For example, the analysis based on weekly data showed financial stocks had the most links, followed by industrial stocks. When daily data and shorter time windows are used, stocks from energy and materials sectors  dominate the linkages. That indicates fast-moving energy and raw materials markets influence stocks in these sectors more than the factors responsible for comovement with the financial and industrial sectors. The results demonstrate the existence of sectoral influences. While the extent of comovement fluctuates, their relative influence shows some  stability. The frequency of the data has a significant influence on the visibility of these sectoral influences. The speed and degree of stock comovements across sectors and across borders  varies across time horizons.

Our findings therefore support findings by, for example, \cite{corsetti} who argue that the level of comovement between markets is changing significantly over time, even when taking into account varying levels of market volatility \cite[see also][]{contagion}. 

\begin{table}[h!]
\begin{center}
\small
\begin{tabular}{ c | c  c  c  c  c }
\hline
 & Nodes & Degree & Clustering & p. Length & Density \\
 &   $n$ & $\langle k \rangle$   & $\langle C \rangle$  & $\langle L \rangle$ &  \\
Window \\
		\hline 
Jul 06 -- Jun 07 & 26 & 4.00 & 0.573 & 1.76 & 0.160 \\
%\hline
Jan 07 -- Dec 07 & 50 & 7.12 & 0.730 & 2.50 & 0.145 \\
%\hline
 Jul 07 -- Jun 08 & 66 & 7.80 & 0.716 & 2.34 & 0.120\\
%\hline
  Jan 08 -- Dec 08 & 80 & 9.50 & 0.734 & 2.81 & 0.120 \\
%\hline
 Jul 08 -- Jun 09 & 84 & 9.77 & 0.732 & 2.75 & 0.118 \\
%\hline
Jan 09 -- Dec 09 & 53 & 10.32 & 0.744 & 2.61 & 0.198 \\
%\hline
Jul 09 -- Jun 10 & 63 & 10.24 & 0.654 & 2.50 & 0.165 \\
%\hline
 Jan 10 -- Dec 10 & 60 & 10.78 & 0.683 & 2.33 & 0.183 \\
%\hline
Jul 10 -- Jun 11 & 42 & 7.35 & 0.686 & 1.93 & 0.179 \\
%\hline
Jan 11 -- Dec 11 & 71 & 13.45 & 0.729 & 2.57 & 0.192 \\
%\hline
 Jul 11 -- Jun 12 & 75 & 13.45 & 0.794 & 2.36 & 0.182 \\
%\hline
Jan 12 -- Dec 12 & 37 & 8.03 & 0.706 & 2.27 & 0.223\\
%\hline
Jul 12 -- Jun 13 & 36 & 6.31 & 0.650 & 1.92 & 0.180\\
\hline
\end{tabular}
\caption{Summary Statistics for Sector-based Networks Calculated Using Dynamical Analysis. For each window we calculate the number of nodes, $n$ , average degree, $ \langle k \rangle$, average clustering coefficient, $\langle C \rangle$, average shortest path for the giant component, $\langle L \rangle$, and density.}
\label{tab:DNstats}
\end{center}
\end{table}

We can also quantify the changing comovement with network statistics. Table \ref{tab:DNstats} illustrates that the number of nodes with statistically significant relationships fluctuates between 26 and 84. The average number of links of each node, the degree, is related to the number of nodes. This relationship is roughly linear but has significant variation. This indicates that the network is not just undergoing size but also structural changes. This is confirmed by the clustering statistics. Clustering measures how often two nodes B and C, which are both connected to node A, will also be connected to each other. All networks show clustering, the highest levels are observed in 2011. This shows that the networks do change along several dimensions and even in times of crisis not everything is always connected to everything else. This is confirmed by the changes in the average path length. Their changes are not only caused by network size changes but also by the contractions and diversions of single components of the network over time. The density describes the fraction of existing versus possible links between nodes. These values are high because of many links between sectors within countries that are rather stable. We see the highest values during the financial crisis, when the network is large and very connected, and in 2012, when the network is smaller and dense. These findings indicate fundamental changes in the network structure over time which, as we will also see in the next Section, cannot be fully explained by the country-level or sector-level influences described above.

\section{Determinants of network structure}\label{sec:ergm}
 
The visualizations of the stock dependencies give the impression of a network structure that is governed by country-wise grouping. In the following, we analyze different possible hypotheses that can explain this structure. Furthermore, we investigate the extent to which sector-related effects can help to explain the observed interdependencies. 

We estimate different exponential random graph models to compare the hypotheses. The ERGM is employed with the unweighted undirected networks on the stock level. Although the estimation technique is rather involved it helps to realize that the results of this estimation can in fact be interpreted like the outcome of a logistic regression. Hence, our dependent variable are the links in our network and we estimate what influence certain variables have on the probability of having a link between pairs of stocks with certain characteristics.

We start with a baseline model, in which the level of links depends on the country and the sector that the stocks belong to. Further, we estimate the additional effect of pairs of stocks that are part of the same country and the same sector. We find that such a baseline structural model explains a good amount of the network structure. All country and sector effects are highly significant. The largest coefficients on the country level are estimated for the U.S., the Netherlands and Hong Kong. On the sector level the effects are largest for the sectors energy, financials and materials (these are effects on connectivity that are not explained by the number of stocks).  The variable for within-country links is highly significant, as table \ref{tab:ergm} shows. Within-sector links are also significant, yet the effect is one order of magnitude smaller. Next, we analyze which additional determinants might improve the overall fit of the model in terms of the model AIC.\footnote{Due to the large number of observations the AIC and the BIC are in this case almost identical. In all cases the null deviance is 10,154,437 and the AIC/BIC are very close to the residual deviance (minus compensation for the number of variables).} 

\begin{sidewaystable}
%\setlength{\tabcolsep}{4pt}
%\footnotesize

    \begin{tabular}{c | c c c c c c c c c c c c }
    \hline
    \hline
			
model & \multicolumn{2}{c}{Base}  & \multicolumn{2}{c}{West vs. East} & \multicolumn{2}{c}{Development} & \multicolumn{2}{c}{Sectoral diff.} & \multicolumn{2}{c}{Sector by sec.} & \multicolumn{2}{c}{Country by co.} 
 \\
 AIC & \multicolumn{2}{c}{5,376,507} & \multicolumn{2}{c}{5,269,102} & \multicolumn{2}{c}{5,276,926} & \multicolumn{2}{c}{5,369,822}  & \multicolumn{2}{c}{5,370,925} & \multicolumn{2}{c}{4,763,994}\\
 nvar & \multicolumn{2}{c}{25} & \multicolumn{2}{c}{26} & \multicolumn{2}{c}{26} & \multicolumn{2}{c}{26}  & \multicolumn{2}{c}{79} & \multicolumn{2}{c}{144}\\
 & coef. & $Z$ & coef. & $Z$ & coef. & $Z$ & coef. & $Z$ & coef. & $Z$  & coef. & $Z$\\
 \hline
\\	
country effects & $\bullet $ & $\bullet $ & $\bullet $ & $\bullet $ & $\bullet $ & $\bullet $ & $\bullet $ & $\bullet $ & $\bullet $ & $\bullet $ & $\bullet $ & $\bullet $ \\
sector effects  & $\bullet $ & $\bullet $ & $\bullet $ & $\bullet $ & $\bullet $ & $\bullet $ & $\bullet $ & $\bullet $ & $\bullet $ & $\bullet $ & $\bullet $ & $\bullet $ \\
match country & 4.6307 & (1064.9) & 4.0338 & (910.0) & 3.9155 & (857.7) & 4.1530 & (580.9) & 4.6215 & (1057.6) \\
match sector & 0.1499 & (47.1)   & 0.1516 & (47.0) & 0.1491 & (46.3) & 0.1388 & (43.6) & & & 0.1707 & (50.0) \\
match region & & & 0.7971 & (328.3)\\
match devmnt. & & & & & 0.8799 & (319.9) \\
sectoral diff. & & & & &  & & -7.1627 & (81.7) \\
interact. sector & & & & & & & & & $\bullet $ & $\bullet$ \\
interact. country & & & & & & & & & & & $\bullet $ & $\bullet$ \\
\hline

 \end{tabular}
 \caption{Estimation Results for Determinants of Network Structure on the Level of Stocks. We show the results for six different models. In each case we report the most important coefficients together with their Z-score. Country-based effects and regional segmentation explain large parts of the observed network structure while sectoral effects are also significant, yet on a smaller scale. }\label{tab:ergm}
 \end{sidewaystable}

In the first model specification, we assume that regional aspects might help to explain the number of links. The simplest assumption is segregation into countries of the "west" versus the "east".\footnote{BRA, ESP, FRA, GBR, USA, CAN, GER vs. CHN, HKG, IND, KOR, SGP, JPN. Alternatively we have grouped the countries into three groups, "America", "Europe", and "Asia", with almost identical explanatory power.} The estimated effect for links within these groups is in fact highly significant and improves the overall fit. Similar results were obtained by \cite{det_inte}. Alternatively, in our second model specification, we group the countries by economic development, i.e. developed versus developing countries.\footnote{AUS, ESP, FRA, GBR, JPN, NLD,  USA, CAN, GER, HKG vs. BRA, CHN, IND, KOR, SGP.} The effect and significance of this variable is similar to the regional grouping.

As a third model specification, we focus on an approach to relate differences in linkages between countries to differences in industrial composition \citep[see, e.g., ][]{griffin}. In our case we can measure the difference between countries by the relative number of firms $n$ in each sector $s$. We calculate a proxy for sectoral differences by calculating the average absolute difference of the fractions in each sector, so that for two countries $i$ and $j$\footnote{Using squared differences leads to almost indistinguishable yet slightly less significant results in the estimation.}
 
\begin{equation}
	sd_{i,j} = <|\frac{n_{s,i}}{N_i}-\frac{n_{s,j}}{N_j}| >
\end{equation}

When estimating a model that employs these sectoral differences as a covariate for the links, we find that it is significant, yet not on the same level as the previous grouping variables. It should be noted that we only capture effects from composition here, overall differences in connectivity between sectors or size effects are already being accounted for. As a robustness check, we have tested if this changes if we omit the country-fixed effects. Yet, in this case the $Z$-score falls to 60.5 and the AIC worsens to 6,681,810.

In two additional model specifications, we investigate two extreme cases, namely how well we explain the network structure when we estimate effects for all sector-to-sector or all country-to-country interactions. The results for the first case show that sector interaction effects on the global level achieve only a very small improvement in AIC compared to the baseline model. Hence, while the levels for different sectors matter, the interaction effects provide only very little additional information. On the other hand, when we estimate interaction effects for all the countries, the improvement in the AIC is very noticeable. This is of course mostly to be expected, given the number of variables and the structure of the data. Nevertheless, it also shows that effects additional to those that we have just tested are present in the data and that most of those manifest itself on the country-to-country level (a further visualization of the sector-by-sector linkages is provided in appendix \ref{sec:appsec}).

\begin{figure}[tb]
\begin{center}
  \includegraphics[width=\textwidth, trim= 0 300 10 300 , clip=true]{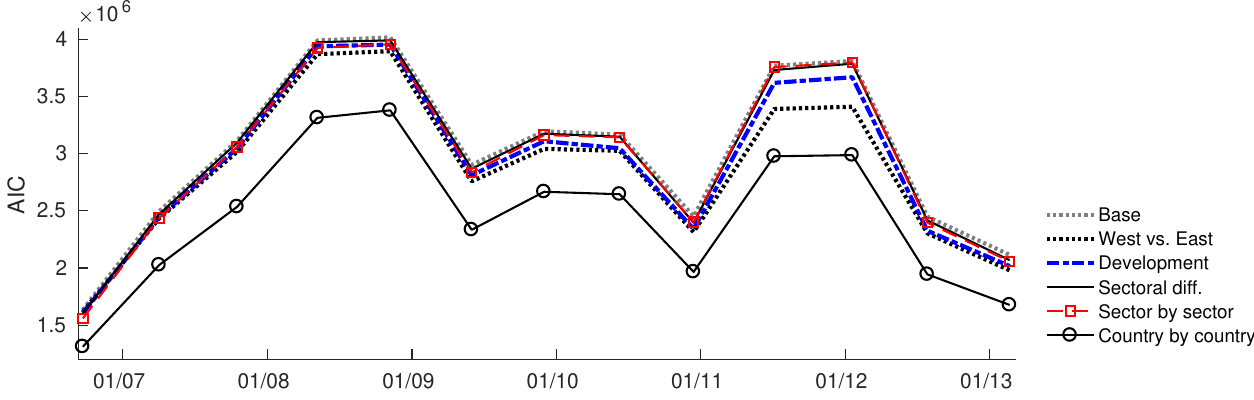}
\end{center}
  \caption{Comparison of Models over Time by out-of-sample Prediction Error. We report the AIC score for the models as specified in table \ref{tab:ergm} but estimate them with the daily data for the 13 time windows as specified in table \ref{tab:DNstats}. The prediction errors move largely in unison yet they vary significantly over time. In all cases the models perform worse in times of crisis. }\label{fig:ergm}
\end{figure}

Finally, we investigate how the performance of these different models changes when we estimate them for the 13 time windows that were discussed in the previous section. The AICs for the models for each time window are presented in Figure \ref{fig:ergm}. First, we observe that the performance of the models mostly changes together across time windows. All models perform worse during the times of the financial crisis of 2008 and during the European debt crisis than in "normal" times. Moreover, in these normal times also all of our models perform much better than in the estimation conducted for the entire time period. The only major shift in relative model performance is then in fact observed at the peak of the European debt crisis.\footnote{Plans for a referendum for Greece to leave the EMU we made in October 2011 yet factually abandoned by the end of 2011.} At this point the models that propose segregation between the western or developed markets and the rest perform noticeably better than the base model. This also confirms our impression from the network visualizations in the last section that hinted at stronger comovement within the Asian countries in 2011. 

These results hint at two things. First, not surprisingly,  we can see that the European debt crisis was less of a global crisis than the crisis of 2008. Secondly, the underlying structural properties of market interdependencies are getting cloaked by different temporary effects in the times of crisis. Hence, very long sample periods do not necessary help in the identification of determinants of comovement. For the case of sectoral effects our results tend to support findings by \cite{griffin} and \cite{dutt}. Effects from sectoral composition do exist, yet their influence is not decisive in explaining the variations in stock interdependencies. We also observed that the network structure shows significant interaction effects on the country-by-country level, while the sector-by-sector level shows only little of such effects. This finding explains why cross-country factors are much easier to identify than sectoral factors and mirrors the findings of \cite{forbchinn} and other factor-model approaches.

\section{Conclusions}

This paper presents an empirical investigation of comovement, as a statistical measure of interconnectedness, across 15 representative stock markets.  We present a framework to measure such statistical interconnectedness on different levels of granularity - the stock, the sector, and the country level. The analysis shows that for  asset markets, a global financial market, in terms of cross-country statistical interconnectedness, exists only within certain limits. We observe a mixture of regional and global sectoral effects, with the balance between the two fluctuating over time. For most of the time, regional segregation remains visible, even though in times of stress markets contract to a unisonous behavior. During such contraction periods, some countries still retain a high level of autonomous behavior.

Previous research has focused on factors that determine the level of stock market comovement for various countries. The results, however, were not always as convincing as similar studies on the synchronization of business cycles. Our findings shed some light on this debate: the fine structure of stock market comovement shows significant time variation. Sectoral effects do exist, but they influence only parts of markets. These effects are also overshadowed by bubbles or collapses such as  the financial crisis. Tracking the sectoral interconnections over time, we find for example a shift from the financial sector, to the materials and energy sectors. A similar effect has been found during the dot-com bubble of 1997-2000 \citep[see also][]{imbs,wagner}. The observation of these shifts help explain why it is difficult to identify determinants for comovement as well es sectoral factors.  %In fact, for the Western markets, country determinants are probably at best secondary effects. 

Our results also hint that the dimensional reduction on the level of sectors preserves
qualitative features of the global stock market, like sectoral shifts, that can be lost in an analysis that is based on market indices. While indices' comovement is high and persistent, stock and sector-based comovement show more heterogeneity, which is not averaged out in the aggregate.

In general, the methodology presented here can be used to quantify statistical interconnectedness among a large number of assets in many markets. We quantify within and between market interconnectedness, and use network theory to present, quantify, and monitor these relationships and how they change in time. The model presented here is a way to approximate asset-specific risks. First-order risks arise from direct interconnections on the asset level. Second-order risks arise from interconnections on the sector level. Both risk channels must be considered when analyzing the risks for an assets based on regional, global or systemic changes.

%While the focus of this paper is the macroscopic features of stock markets, future work could apply and extend this framework to portfolio decisions for specific stocks. This would necessitate the estimation of covariances at a much higher frequency, which could be achieved by using high-frequency data and/or the estimation of dynamic correlations similar to those of multivariate GARCH models. Our approach could also be extended to include macroeconomic variables and data on bilateral relationships as determinants of asset comovement. This would connect our approach to the factor-model literature.

\section*{Acknowledgments}
\noindent The authors are grateful for discussions at the Kiel Institute's staff seminar, the staff seminar at Kiel University, the Center for Polymer Studies at Boston University, and the participants at the NBER summer institute and the conferences of the EEA, CEF, Bank of Finland and WEHIA. MR is grateful for partial funding of this research from the European Union Seventh Framework
Programme (FP7/2007-2013) under grant agreement no. 619255. 
The article is also part of a research initiative launched by the Leibniz Community.
This paper benefited from helpful comments by Mark Flood, Shlomo Havlin, H. Eugene Stanley, Stacey Schreft, Charles Taylor, Julie Vorman, and Peyton Young.

\section*{References}
\small
\bibliographystyle{elsarticle-harv} 
\bibliography{bib}

\begin{thebibliography}{49}
\expandafter\ifx\csname natexlab\endcsname\relax\def\natexlab#1{#1}\fi
\expandafter\ifx\csname url\endcsname\relax
  \def\url#1{\texttt{#1}}\fi
\expandafter\ifx\csname urlprefix\endcsname\relax\def\urlprefix{URL }\fi

\bibitem[{Aielli(2013)}]{aielli}
Aielli, G.~P., 2013. Dynamic conditional correlation: On properties and
  estimation. Journal of Business \& Economic Statistics 31~(3), 282--299.

\bibitem[{Barberis et~al.(2005)Barberis, Shleifer, and Wurgler}]{barberis}
Barberis, N., Shleifer, A., Wurgler, J., 2005. Comovement. Journal of Financial
  Economics 75~(2), 283--317.

\bibitem[{Baur and Jung(2006)}]{baur}
Baur, D., Jung, R.~C., 2006. Return and volatility linkages between the {US}
  and the {German} stock market. Journal of International Money and Finance 25,
  598--613.

\bibitem[{Bekaert et~al.(2011)Bekaert, Campbell, Lundblad, and
  Siegel}]{segm_equity}
Bekaert, G., Campbell, R.~H., Lundblad, C.~T., Siegel, S., 2011. What segments
  equity markets? The Review of Financial Studies 24~(12), 3841--3890.

\bibitem[{Bekaert et~al.(2009)Bekaert, Hodrick, and Zhang}]{bekhod}
Bekaert, G., Hodrick, R.~J., Zhang, X., 2009. International stock return
  comovement. Journal of Finance 64~(6), 2591--2626.

\bibitem[{Billio et~al.(2012)Billio, Getmansky, Lo, and Pelizzon}]{Billio2012}
Billio, M., Getmansky, M., Lo, A.~W., Pelizzon, L., 2012. Econometric measures
  of connectedness and systemic risk in the finance and insurance sectors.
  Journal of Financial Economics 104~(3), 535--559.

\bibitem[{Bollerslev(1986)}]{Boll}
Bollerslev, T., 1986. Generalized autoregressive conditional
  heteroskedasticity. Journal of Econometrics 31, 307--327.

\bibitem[{Bracker et~al.(1999)Bracker, Docking, and Koch}]{det_inte}
Bracker, K., Docking, D.~S., Koch, P.~D., 1999. Economic determinants of
  evolution in international stock market integration. Journal of Empirical
  Finance 6, 1--27.

\bibitem[{Brooks and Del~Negro(2004)}]{brooks}
Brooks, R., Del~Negro, M., 2004. The rise in comovement across national stock
  markets: Market integration or {IT} bubble? Journal of Empirical Finance
  11~(5), 659--680.

\bibitem[{Buch et~al.(2005)Buch, Doepke, and Pierdzioch}]{opencycle}
Buch, C.~M., Doepke, J., Pierdzioch, C., 2005. Financial openness and business
  cycle volatility. Journal of International Money and Finance 24, 744--765.

\bibitem[{Chen et~al.(2010)Chen, Jiang, Li, and Sim}]{china}
Chen, Z., Jiang, H., Li, D., Sim, A.~B., 2010. Regulation change and volatility
  spillovers: evidence from china's stock markets. Emerging markets finance and
  trade 46~(6), 140--157.

\bibitem[{Christensen et~al.(2010)Christensen, Kinnebrock, and
  Podolskij}]{christensen}
Christensen, K., Kinnebrock, S., Podolskij, M., 2010. Pre-averaging estimators
  of the ex-post covariance matrix in noisy diffusion models with
  non-synchronous data. Journal of Econometrics 159~(1), 116--133.

\bibitem[{Claessens and Kose(2018)}]{claessens2018frontiers}
Claessens, S., Kose, M.~A., 2018. Frontiers of macrofinancial linkages. BIS
  Papers 95.

\bibitem[{Corsetti et~al.(2005)Corsetti, Pericoli, and Sbracia}]{corsetti}
Corsetti, G., Pericoli, M., Sbracia, M., 2005. '{Some} contagion, some
  interdependence': {More} pitfalls in tests of financial contagion. Journal of
  International Money and Finance 24, 1177--1199.

\bibitem[{Diebold and Yilmaz(2009)}]{yilmazvar}
Diebold, F.~X., Yilmaz, K., 2009. Measuring financial asset return and
  volatility spillovers, with application to global equity markets. Economic
  Journal, 158--171.

\bibitem[{Diebold and Yilmaz(2014)}]{diebold2014network}
Diebold, F.~X., Yilmaz, K., 2014. On the network topology of variance
  decompositions: Measuring the connectedness of financial firms. Journal of
  Econometrics 182~(1), 119--134.

\bibitem[{Dutt and Mihov(2013)}]{dutt}
Dutt, P., Mihov, I., 2013. Stock market comovement and industrial structure.
  Journal of Money, Credit and Banking 45~(5), 891--911.

\bibitem[{Engle(2002)}]{ddcg}
Engle, R.~F., 2002. Dynamic conditional correlation: A simple class of
  multivariate generalized autoregressive conditional heteroskedasticity
  models. Journal of Business \& Economic Statistics 20~(3), 339--350.

\bibitem[{Engle et~al.(1990)Engle, Ito, and Lin}]{meteor}
Engle, R.~F., Ito, T., Lin, W.-L., 1990. Meteor showers or heat waves?
  {Heteroskedastic} intra-daily volatility in the foreign exchange market.
  Econometrica 58.

\bibitem[{Engle et~al.(2019)Engle, Ledoit, and Wolf}]{large_dcc2}
Engle, R.~F., Ledoit, O., Wolf, M., 2019. Large dynamic covariance matrices.
  Journal of Business \& Economic Statistics 37.

\bibitem[{Forbes and Chinn(2004)}]{forbchinn}
Forbes, K.~J., Chinn, M.~D., 2004. A decomposition of global linkages in
  financial markets over time. The Review of Economics and Statistics 86~(3),
  705--722.

\bibitem[{Forbes and Rigobon(2002)}]{contagion}
Forbes, K.~J., Rigobon, R., 2002. No contagion, only interdependence: measuring
  stock market comovements. The Journal of Finance 57~(5), 2223--2261.

\bibitem[{Fratzscher(2012)}]{fratzscher}
Fratzscher, M., 2012. Capital flows, push versus pull factors and the global
  financial crisis. Journal of International Economics 88, 341--356.

\bibitem[{Fry et~al.(2010)Fry, Martin, and Tang}]{fry}
Fry, R., Martin, V.~L., Tang, C., 2010. A new class of tests of contagion with
  applications. Journal of Business \& Economic Statistics 28~(3), 423--437.

\bibitem[{Gai et~al.(2011)Gai, Haldane, and Kapadia}]{gaiandandy}
Gai, P., Haldane, A., Kapadia, S., 2011. Complexity, concentration and
  contagion. Journal of Monetary Economics 58~(5), 145--136.

\bibitem[{Glasserman and Young(2016)}]{glasserman2016contagion}
Glasserman, P., Young, H.~P., 2016. Contagion in financial networks. Journal of
  Economic Literature 54~(3), 779--831.

\bibitem[{Gopikrishnan et~al.(2001)Gopikrishnan, Rosenow, Plerou, and
  Stanley}]{Gopikrishnan}
Gopikrishnan, P., Rosenow, B., Plerou, V., Stanley, H.~E., 2001. Quantifying
  and interpreting collective behavior in financial markets. Physical Review E
  64~(3), 035106.

\bibitem[{Green and Hwang(2009)}]{green}
Green, T.~C., Hwang, B.-H., 2009. Price-based return comovement. Journal of
  Financial Economics 93~(1), 37--50.

\bibitem[{Greenwood-Nimmo et~al.(2015)Greenwood-Nimmo, Nguyen, and
  Shin}]{largevar}
Greenwood-Nimmo, M., Nguyen, V.~H., Shin, Y., 2015. Measuring the connectedness
  of the global economy. Melbourne Institute Working Paper Series No. 7/15.

\bibitem[{Griffin and Karolyi(1998)}]{griffin}
Griffin, J.~M., Karolyi, G.~A., 1998. Another look at the role of the
  industrial structure of markets for international diversification strategies.
  Journal of Financial Economics 50, 351--373.

\bibitem[{Hamao et~al.(1990)Hamao, Masulis, and Ng}]{hamao1990correlations}
Hamao, Y., Masulis, R.~W., Ng, V., 1990. Correlations in price changes and
  volatility across international stock markets. The Review of Financial
  Studies 3~(2), 281--307.

\bibitem[{Hansen and Lunde(1998)}]{hanslund}
Hansen, P.~R., Lunde, A., 1998. A forecast comparison of volatility models:
  Does anything beat a {GARCH(1,1)}? Journal of Applied Econometrics 20,
  873--889.

\bibitem[{Hayashi and Yoshida(2008)}]{hayashi}
Hayashi, T., Yoshida, N., 2008. Asymptotic normality of a covariance estimator
  for nonsynchronously observed diffusion processes. Annals of the Institute of
  Statistical Mathematics 60~(2), 367--406.

\bibitem[{Hu(2005)}]{hu}
Hu, Y., 2005. Efficient, high-quality force-directed graph drawing. Mathematica
  Journal 10~(1), 37--71.

\bibitem[{Imbs(2004)}]{imbs}
Imbs, J., 2004. Trade, finance, specialization, and synchronization. The Review
  of Economics and Statistics 86~(3), 723--734.

\bibitem[{Kaminsky and Reinhart(2000)}]{kaminsky}
Kaminsky, G., Reinhart, C., 2000. On crises, contagion, and confusion. Journal
  of International Economics 51~(1), 145--136.

\bibitem[{Kenett et~al.(2012{\natexlab{a}})Kenett, Raddant, Lux, and
  Ben-Jacob}]{plos}
Kenett, D.~Y., Raddant, M., Lux, T., Ben-Jacob, E., 2012{\natexlab{a}}.
  Evolvement of uniformity and volatility in the stressed global financial
  village. PloS one 7~(2), e31144.

\bibitem[{Kenett et~al.(2012{\natexlab{b}})Kenett, Raddant, Zatlavi, Lux, and
  Ben-Jacob}]{kenett2012correlations}
Kenett, D.~Y., Raddant, M., Zatlavi, L., Lux, T., Ben-Jacob, E.,
  2012{\natexlab{b}}. Correlations and dependencies in the global financial
  village. International Journal of Modern Physics 16, 13--28.

\bibitem[{Lange et~al.(1989)Lange, Little, and Taylor}]{robust}
Lange, K.~L., Little, R.~J., Taylor, J.~M., 1989. Robust statistical modeling
  using the t distribution. Journal of the American Statistical Association
  84~(408), 881--896.

\bibitem[{Lincoln and Gerlach(2004)}]{lincbook}
Lincoln, J.~R., Gerlach, M.~L., 2004. {Japan's} Network Economy. Cambridge U.
  P.

\bibitem[{Lusher et~al.(2011)Lusher, Koskinen, and Robins}]{ergmbook}
Lusher, D., Koskinen, J., Robins, G., 2011. Exponential Random Graph Models for
  Social Networks. Cambridge University Press.

\bibitem[{Martens and Poon(2001)}]{martens}
Martens, M., Poon, S.-H., 2001. Returns synchronization and daily correlation
  dynamics between international stock markets. Journal of Banking \& Finance
  25~(10), 1805--1827.

\bibitem[{Martinez-Jaramillo et~al.(2019)Martinez-Jaramillo, Carmona, and
  Kenett}]{martinez2019interconnectedness}
Martinez-Jaramillo, S., Carmona, C.~U., Kenett, D.~Y., 2019. Interconnectedness
  and financial stability. Journal of Risk Management in Financial Institutions
  12~(2), 168--183.

\bibitem[{Newman(2010)}]{newman}
Newman, M.~E., 2010. Networks: {An Introduction}. Oxford University Press.

\bibitem[{Raddant and Wagner(2017)}]{wagner}
Raddant, M., Wagner, F., 2017. Transitions in the stock markets of the {US},
  {UK}, and {Germany}. Quantitative Finance 17~(2), 289--297.

\bibitem[{Rigobon(2003)}]{rigobon_jie}
Rigobon, R., 2003. On the measurement of the international propagation of
  shocks: is the transmission stable? Journal of International Economics 61,
  261--283.

\bibitem[{Song et~al.(2011)Song, Tumminello, Zhou, and Mantegna}]{structure}
Song, D.-M., Tumminello, M., Zhou, W.-X., Mantegna, R.~N., 2011. Evolution of
  worldwide stock markets, correlation structure, and correlation-based graphs.
  Physical Review E 84~(2), 026108.

\bibitem[{Strauss and Ikeda(1990)}]{strauss}
Strauss, D., Ikeda, M., 1990. Pseudolikelihood estimation for social networks.
  Journal of the American Statistical Association 95, 204--212.

\bibitem[{Summer(2013)}]{summer2013financial}
Summer, M., 2013. Financial contagion and network analysis. Annu. Rev. Financ.
  Econ. 5~(1), 277--297.

\end{thebibliography}
\newpage
%\section*{Figures and Tables}

\appendix
%\section*{Appendix}
\section{Statistical Properties of the Data}
\setcounter{figure}{0}
\setcounter{table}{0}

\begin{figure}[h!]
\begin{center}
\includegraphics[width=0.7\textwidth]{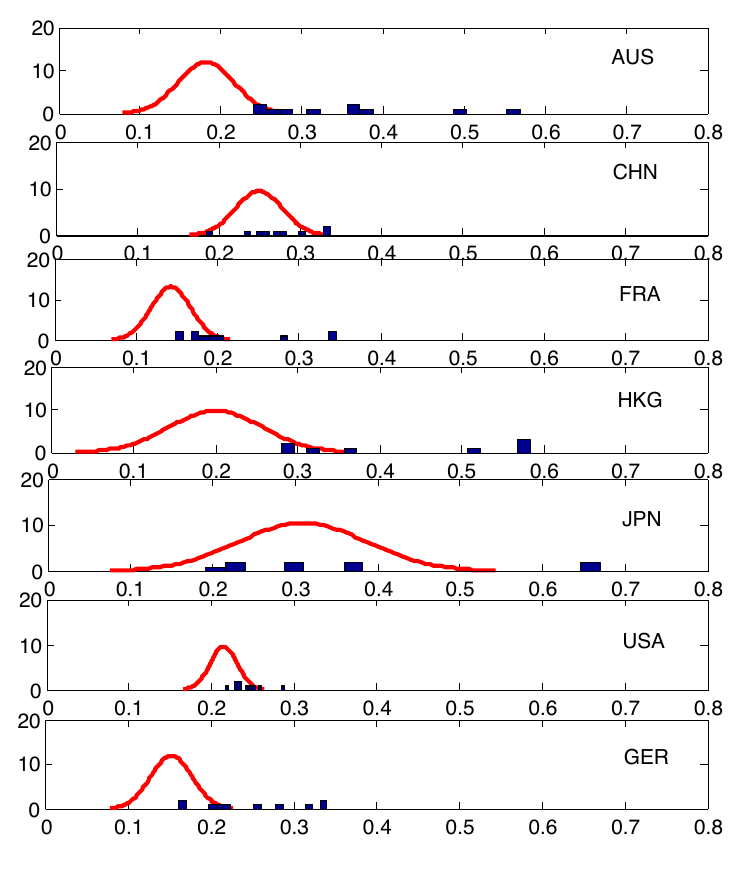}
\end{center}
   \caption{\label{fig:sec_corr} Distribution of Between-Sector Correlations of Stocks (red curve) versus Average Correlations Within the Same Sector (bars) for Seven Countries. The correlation of stocks within sectors is mostly significantly higher than the average correlation and show higher dispersion. The remaining eight markets behave similarly. Only China and most of the Japanese market demonstrated a different behavior.}
\end{figure}

For most markets, the correlation of stocks within a sector is much higher than the overall correlation. Figure \ref{fig:sec_corr} shows details for seven countries. The overall correlations are shown as a fitted normal distribution and the single sector averages are shown as histogram bars. The dispersion of measured correlations is smallest for the United States and highest for Japan. The latter, together with China and Korea, are the only countries where the sector effects are weak.

%Figure \ref{fig:acf} illustrates the ACF of the raw and the filtered returns. For all countries the GARCH(1,1) produces filtered returns without significant auto-correlation. 
%
%\begin{figure}[h!]
%\begin{center}
%\includegraphics[width=\textwidth]{autocorr.pdf}
   %\caption{\label{fig:acf} Average Auto-correlation Functions for the Absolute Raw (left) and Filtered Returns (right) up to Lag 50 for Seven Countries.  Although the markets show different levels of autocorrelation and different speed of decay, the GARCH(1,1) is able to produce de-garched time series which sufficient (insignificant) levels of auto-correlation. The results of the other eight markets are very similar. Sources: Standard \& Poor's Compustat, Thompson Reuters Datastream, author's analysis.}
	%\end{center}
%\end{figure}
\newpage

\begin{figure}
\begin{center}
\includegraphics[width=0.95\textwidth]{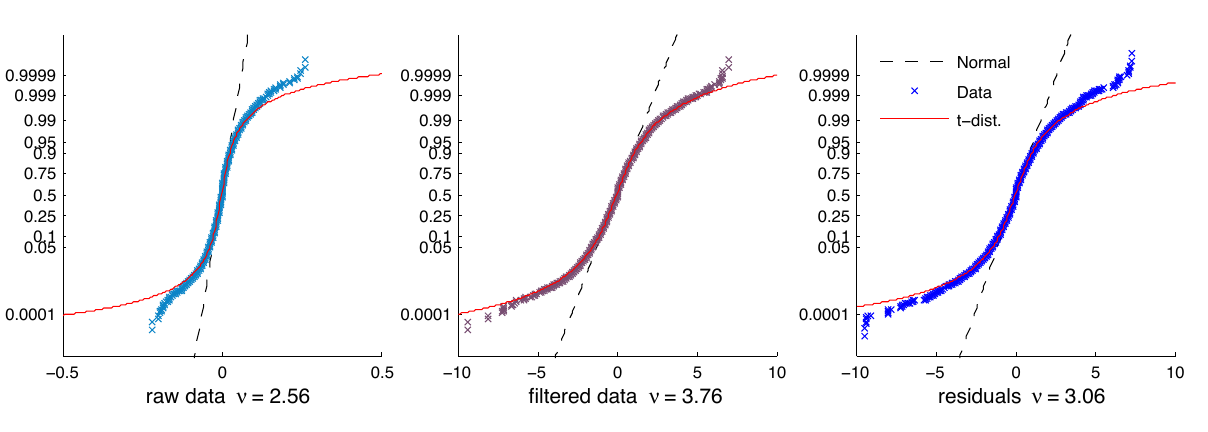}
   \caption{\label{fig:distr} Distributions of the Raw Returns (left panel), Filtered Returns (middle panel) and Residuals (right panel) and Fitted t-distributions. Sub-sample of roughly 20,000 observations. The degrees of freedom for the fitted t-distribution are given below each plot.}
	\end{center}
\end{figure}

Figure \ref{fig:distr} sheds some light on the distributional properties of the raw and filtered data. Not surprisingly the raw returns have very pronounced tails. After normalizing everything with the estimated volatilities the filtered returns are rather well described by the fitted t-distribution (middle panel). The analysis of the residuals (right panel) reveals few observations in the tails that are slightly off, but these are already present in the raw data. Given the large number of observations, these do not pose any problem.

\clearpage
\section{Robustness Analysis}\label{app:dcc}
\setcounter{figure}{0}
The procedure for the estimation of the dependencies is a two-step process. First we remove the volatility changes from our data. Second is the pairwise robust regression. For both steps we have performed robustness tests. 
A test for the first step is to check whether the estimated dependencies differ from the results of a standard multivariate GARCH model. This exercise should ensure that our filtering method is not biased when compared to multivariate methods. Figures \ref{fig:dcc1} and \ref{fig:dcc2} show the comparison of our results with those of the DCC model, estimated with subsamples of the data.
%
%To compare the correlations for the entire time series as well as for a rolling window, we will estimate the DCC model, which will deliver a time dependent correlation matrix. To limit the computation time we choose a sub-sample of 77 randomly selected stocks and estimate the DDC pairwise. For a detailed discussion of the model the reader is referred to \cite{bauwens} and \cite{ddcg}. In brief, within a multivariate GARCH model one tries to find a matrix of conditional variances denoted $H_t$. In the specification of the DDC model by Engle and Sheppard, it is assumed that the correlations between the assets are time varying, such that (in the $p=q=1$ case)
%
%\begin{eqnarray}
%H_t = D_t R_t D_t,
%\end{eqnarray}
%where
%\begin{eqnarray}
%D_t = diag(h_{11,t}^{1/2},\hdots,h_{NN,t}^{1/2}),
%\end{eqnarray}
%with the $h_t$s as defined before and
%\begin{eqnarray}
%R_t = diag(q_{11,t}^{1/2},\hdots,q_{NN,t}^{1/2}) Q_t diag(q_{11,t}^{1/2},\hdots,q_{NN,t}^{1/2}),
%\end{eqnarray}
%
\begin{figure}[h]
\begin{center}
\includegraphics[width=0.9\textwidth]{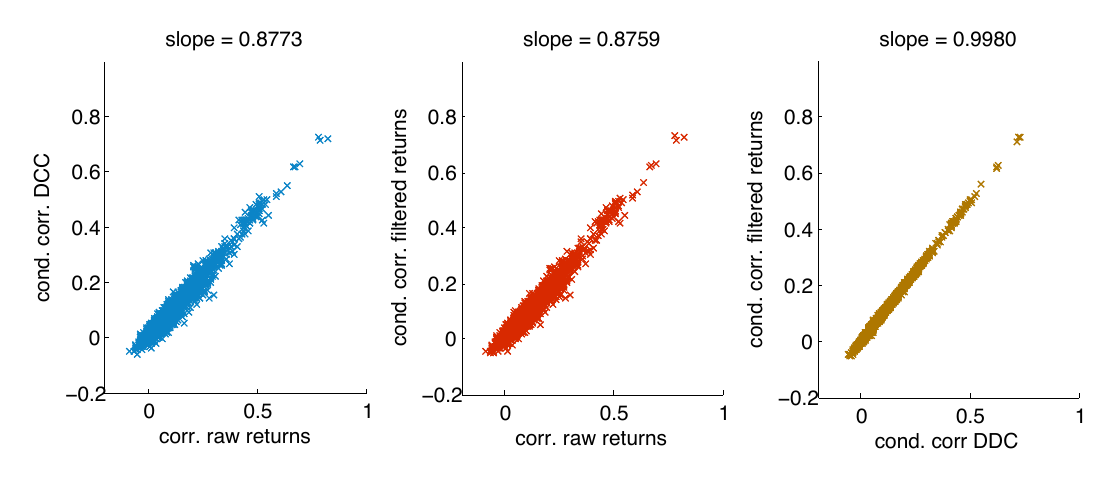}
\caption{\label{fig:dcc1}Comparison of Correlations. The left panel shows the correlation of raw returns vs. the correlation conditional on the volatility estimated by the DCC model. The middle panel shows the same for the correlation vs. the correlation of the GARCH filtered returns. The right panel compares the correlation coefficients from the simple GARCH and DCC model. Both models correct the correlation to about 87 percent of the one that one would get from the raw returns.}
	\end{center}
\end{figure}
%%
%\noindent where
%\begin{eqnarray}
%Q_t=(1-\alpha-\beta)\bar{Q}+\alpha u_{t-1}u^,_{t-1}+\beta Q_{t-1},
%\end{eqnarray}
%where $\alpha$ and $\beta$ are parameters, $\bar{Q}$ is the unconditional variance of $u$, and $u_{it}=\varepsilon_{it}/\sqrt{h_{iit}}$.
%
%
%The estimated volatilities from the univarite GARCH and the DCC (the BEKK model delivers almost identical results) are very similar and hence also the correlation coefficients are almost identical, at least when one averages over the entire time period. The left and middle panel in Figure~\ref{fig:dcc1} both show a scatter plot of the correlation coefficients of the time series of raw returns versus the time series normalized by the estimated volatility. Both correct the calculated correlation coefficient downwards by the same amount. The right panel shows a scatter plot of the correlation coefficients derived from the DCC model versus those from our univariate filtering. 
%
\begin{figure}[ht!]
\begin{center}
\includegraphics[width=0.7\textwidth, trim= 72 218 90 210, clip=true]{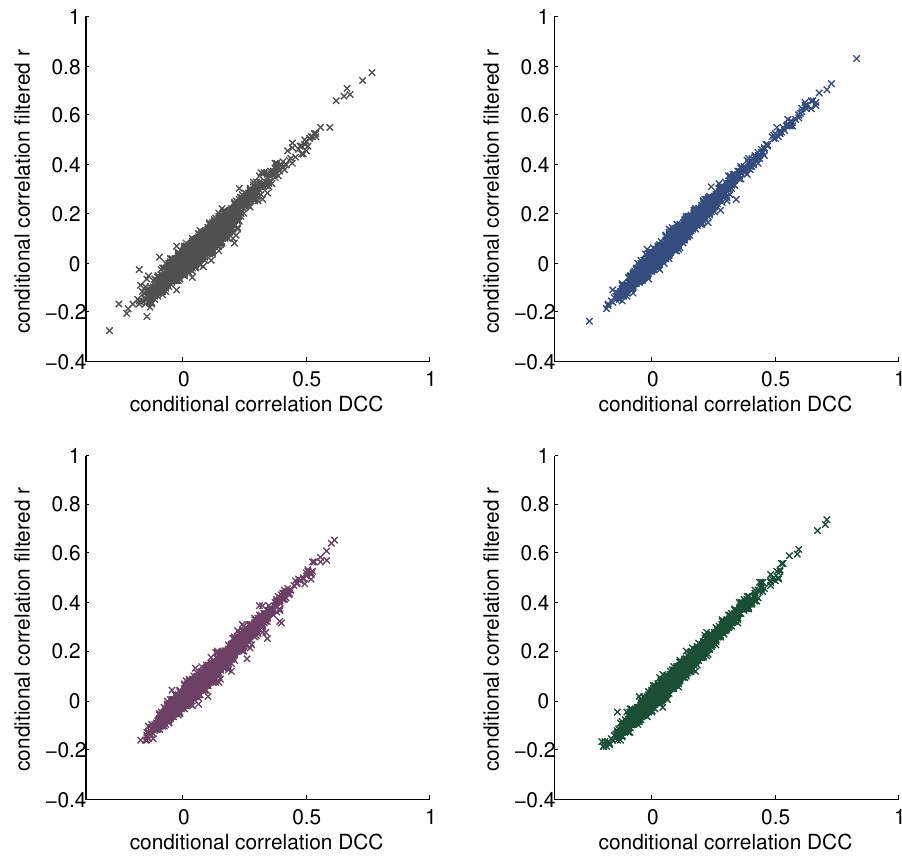}
\caption{\label{fig:dcc2}Scatter Plots of the Correlations of the Filtered Returns versus the Average of the DCC Model Correlation for Four Random Time Windows.}
	\end{center}
\end{figure}
%
%These results also hold, on average, when looking at the averaged dynamic correlation coefficient for the 13 time windows in our analysis. Figure \ref{fig:dcc2} shows scatter plots of correlation coefficients for filtered returns time series (univariate vs. DCC model) for four of the 13 time windows. The plots show slightly greater variation than the plots for the entire time period.
%
%The results also indicate it is possible to look at correlation networks on a much smaller time scale than in the remainder of the paper if using the dynamic correlation coefficients from the DCC model. This would, however, require a massive computer cluster using parallel computation.
%
For the second step of our estimation we rely on the results of a robust estimation and the resulting p-values. For this step it is necessary to examine the distributions of the p-values. We checked if the estimation is sensitive to the large sample size, hence, if the large amount of pairwise comparisons might lead to spurious correlations.

\begin{figure}[ht!]
\begin{center}
\includegraphics[width=0.7\textwidth, trim= 120 280 120 240, clip=true]{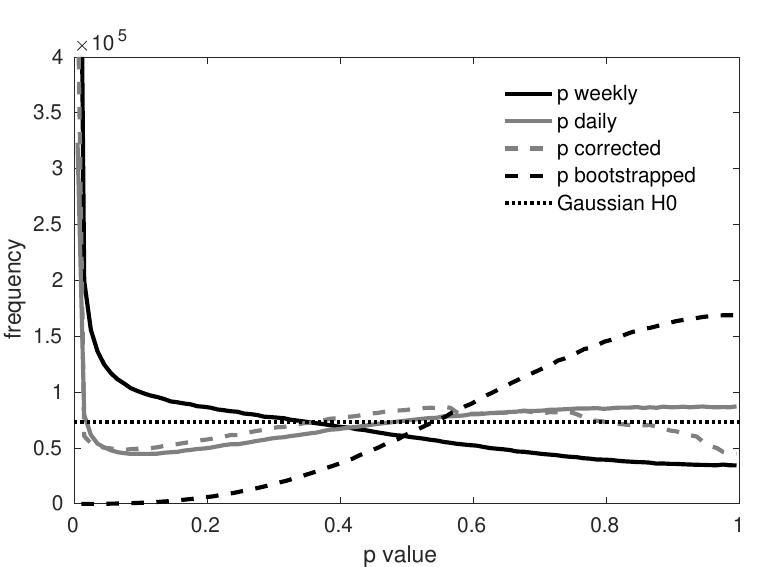}
\caption{\label{fig:pvals} Distributions of p-values. The plot shows the distributions of the p-values of the  weekly (black line) and one random (yet representative) example of the daily regressions (gray line). Both show a peak near zero and a distribution close to a uniform  for higher values. The weekly regression tends towards higher significance, which can be expected since weekly stock returns tend to be more closely correlated. The results from the timing-corrected estimation are shown as the broken gray line, with the expected effects of the correction. In order to check for possible problems related to a multiple comparison problem we have also performed a bootstrapping exercise with permutations of the weekly returns. The distributions of the p-values of this simulation are shown by the broken black line. It reveals that an inflation of false positives is rather improbable since the robust regression does not assign significant amounts of false positive results, especially $p<0.1$, to these uncorrelated time series. Hence, the robust regression is in fact robust against uncorrelated t-distributed noise, other than the known problems with Gaussian estimates, which expected p-values are shown by the black dotted line. (The figure is clipped at the top, the max. for p weekly is around $8.5 x 10^5$, $6.5 x 10^5$ for p corrected, $ N_{obs} = 7.3 x 10^7 $, $ N_{bin} = 100 $.)}
	\end{center}
\end{figure}

Figure \ref{fig:pvals} shows that the distributions of p-values have the expected shape. The number of observed significant interdependencies is far higher than the (conservative) Gaussian null-model. Most connected stocks are significant at a high level, mostly 99\%. The interdependencies of pairs of stocks from different markets reach values just below that, often around 95\%.

As a further check for possible noise in the estimation results we generate permutations of our data (the de-garched returns), such that the distributions of all the time series remain the same but that they resemble uncorrelated noise otherwise. We then repeat the exact same pairwise estimation and plot the distribution of observed p-values. The plot shows that for the p-values used in this analysis false positives are extremely unlikely. 

\clearpage
\section{Relationship of Covariances and Correlation of De-garched Returns}\label{app:corr}

\begin{figure}[hb!]
\begin{center}
\includegraphics[width= \textwidth]{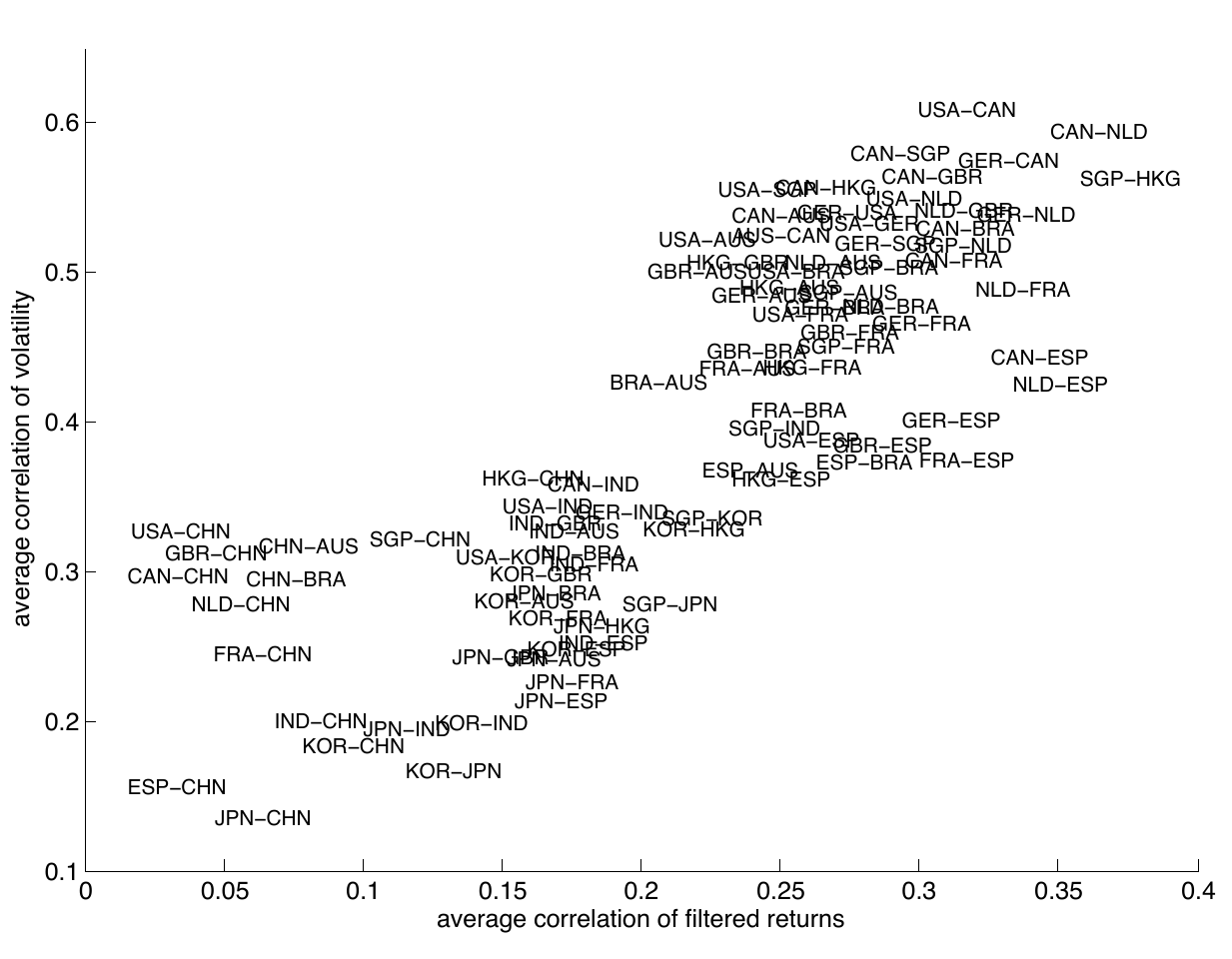}
   \caption{Correlation of Returns and Correlation of Volatility.  We compare the averages of the correlations of the estimated (GARCH) volatilities and the averages of the correlation of the correlation of the filtered returns. Hence, each plotted label represents the averages of the correlation of all stocks in the first with all the stocks in the second country. Although there is a clear positive relationship between correlation of volatility and correlation of filtered returns, some country specific differences in the ratio of the two measures can be observed. The ration is above average for pairs of countries that involve China, and the ration is below average for many pairs that involve European countries.}\label{fig:volret}
	\end{center}
\end{figure}

Throughout the previous analysis, we have used the estimated volatility from the GARCH model to normalize the returns. It can also be useful to consider the correlation of this estimated volatilizes of stocks and check if these are simply proportional to the correlation of normalized stock returns or not. Interestingly, when we average the volatility on a country to country level there is some structure in the behavior of these two measures. Figure \ref{fig:volret} presents a scatter plot of the average correlations of filtered returns versus the average correlations of estimated volatility. 

At a first glance, there seems to be a lot of noise around some imaginary positive-sloped line, but the printed labels reveal some structure. China's stocks are, as discussed above, only weakly correlated with those of the rest of the world, but the correlation of volatility is relatively large, making most of the CHN labels appear above others in the left part of the figure. This is also true for many distant countries, which appear on the top edge of the scatter cloud in the right half. On the other end of the spectrum, we have pairs of mainly European countries, where the ratio of volatility- to returns-correlation is low, which appear a bit below the bulk of the scatter cloud in the right half of the figure. Hence, comovement cannot be synonymously analyzed by either volatility or returns comovement. For China, it seems that financial market restrictions can limit volatility spillovers much less than comovement in returns. For the European markets, we see high levels of comovement in returns but relatively less volatility spillovers than for distant countries.

\clearpage
\section{Dependencies on the Stock and Country Level}
\setcounter{figure}{0}

\begin{figure}[hb!]
\begin{center}
\includegraphics[width=0.85\textwidth]{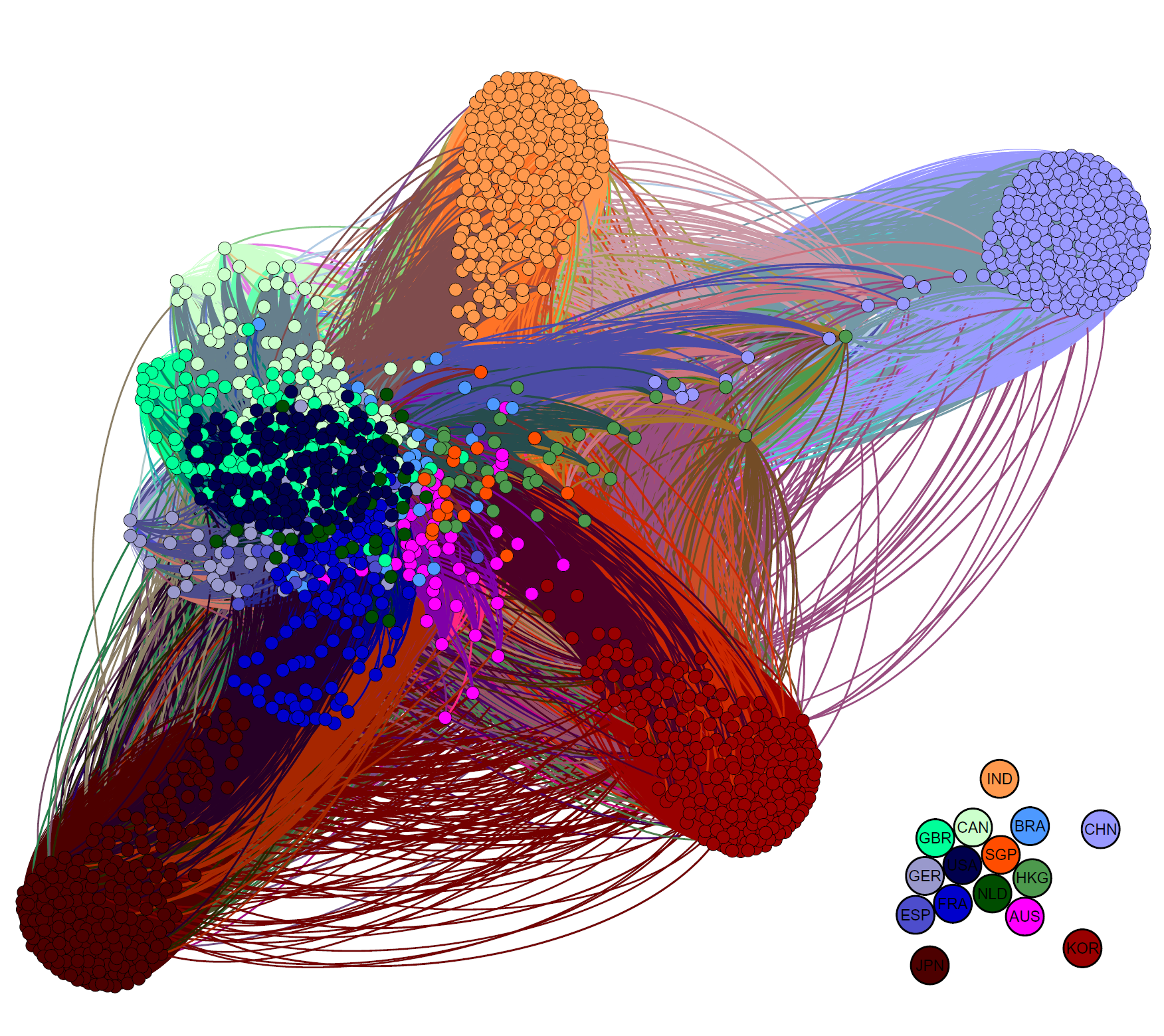}
   \caption{\label{fig:weekly} Network Representation Based on the Estimated Dependencies from Weekly Data.  The figure shows a visualization of the network of stocks where the weighted links correspond to stock dependencies that at least satisfy the 90 percent confidence interval. Stocks from the same country have the same color and a legend of the color coding is at the bottom right. Stocks of most countries form a mixed cluster in the center of the figure. Regional structures and parts of national stock markets are at the periphery of the mixed cluster. Stocks from India, China, Japan, and South Korea are not part of the central cluster and these countries show different levels of connectedness. While parts of the Japanese market seem to form a bridge toward the center, the connections of Chinese stocks are weaker and less diverse. The visualization was performed in Gephi using the Yifan Hu multilevel algorithm.}
	\end{center}
\end{figure}

In Figure \ref{fig:weekly} we show the network representation based on the estimated dependencies from the weekly data, on a stock-stock interaction level. Within the  densely-connected core, most sectors of the U.S. stock market are in the middle. They connect with sectors from European stock markets on the left, with the Asian markets at the top, and with markets from the Americas on the bottomright. This layout is the result of an optimization algorithm and other algorithms deliver similar qualitative results.

\begin{figure}
  \begin{center}
    \subfigure[Jan 07 -- Dec 07]{\includegraphics[width=0.35\textwidth, trim= 0 130 0 130, clip=true]{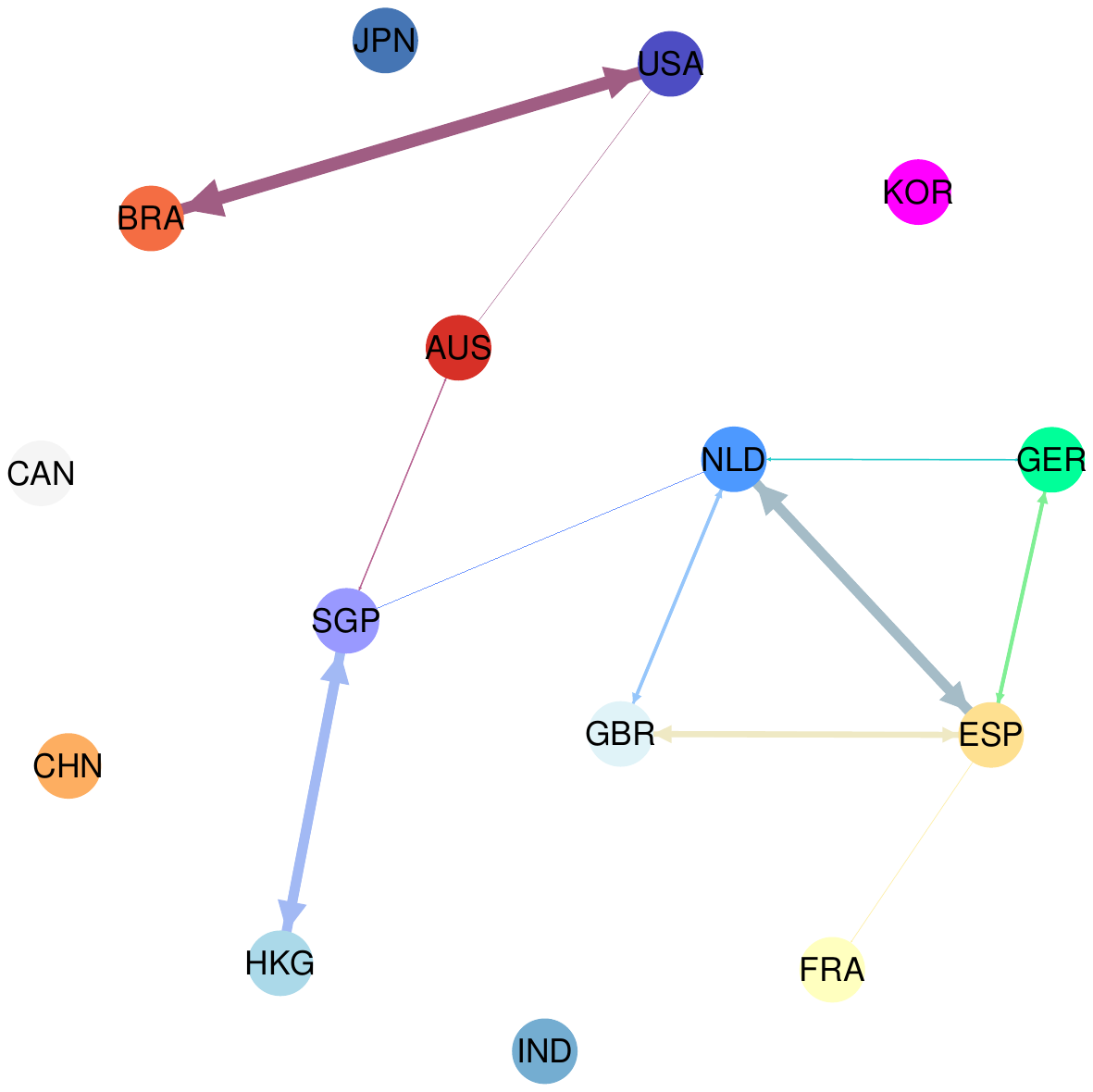}}
		\hspace{0.5cm}
    \subfigure[Jul 08 -- Jun 09]{\includegraphics[width=0.35\textwidth, trim= 0 130 0 130, clip=true]{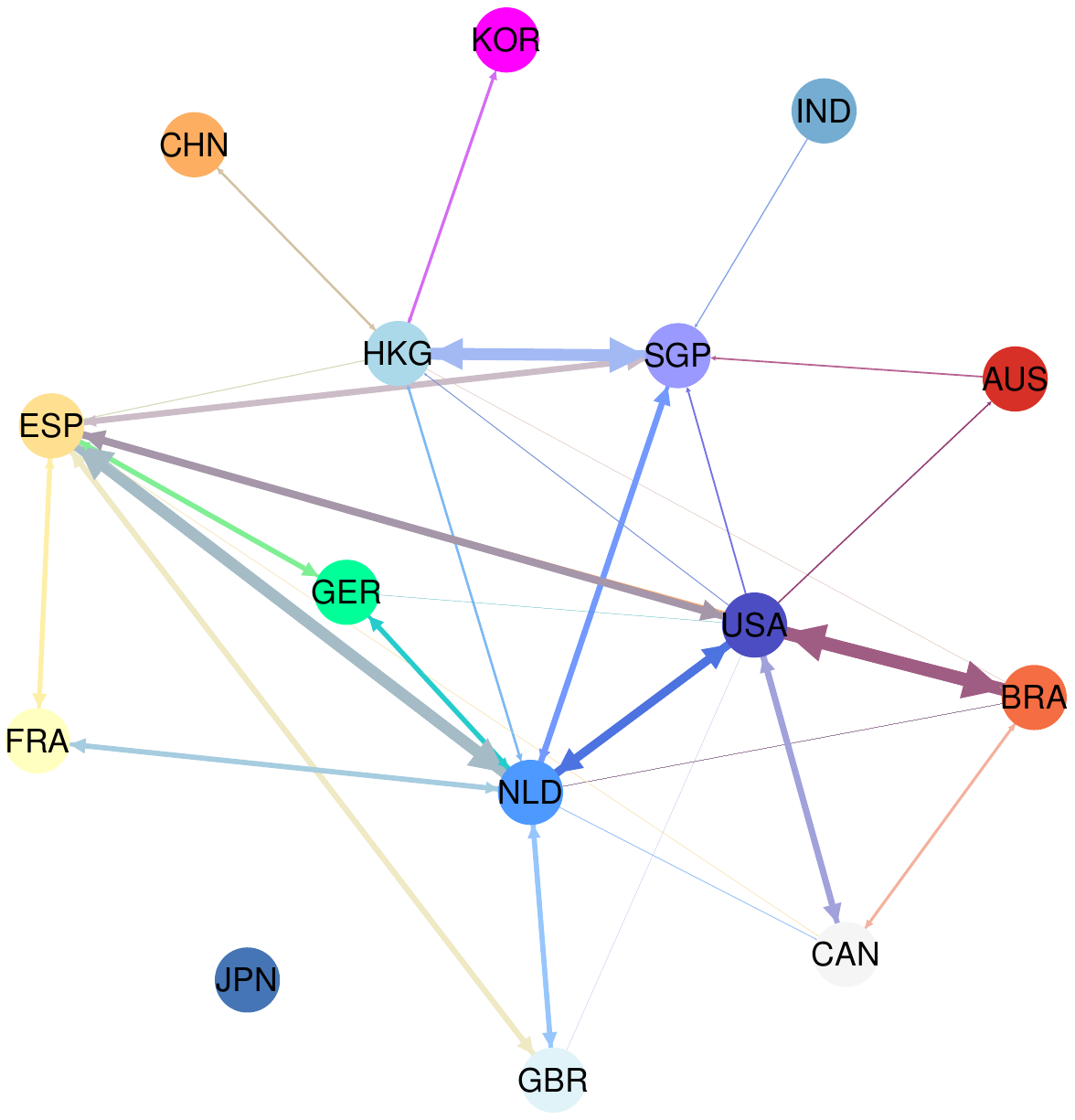}}
    \subfigure[Jan 11 -- Dec 11]{\includegraphics[width=0.35\textwidth, trim= 0 130 0 120, clip=true]{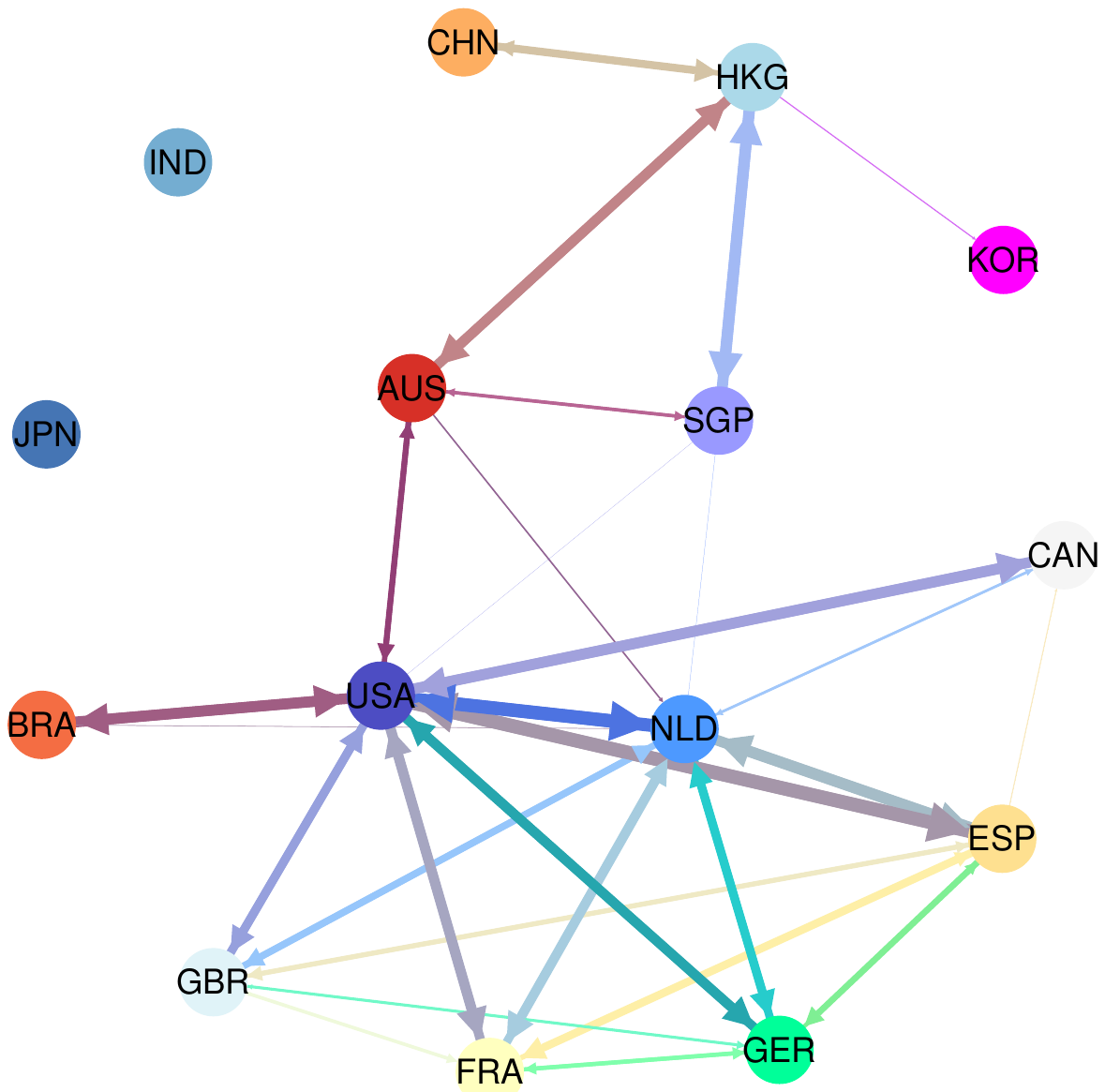}}
		\hspace{0.5cm}
    \subfigure[Jul 12 -- Jun 13]{\includegraphics[width=0.35\textwidth, trim= 0 130 0 120, clip=true]{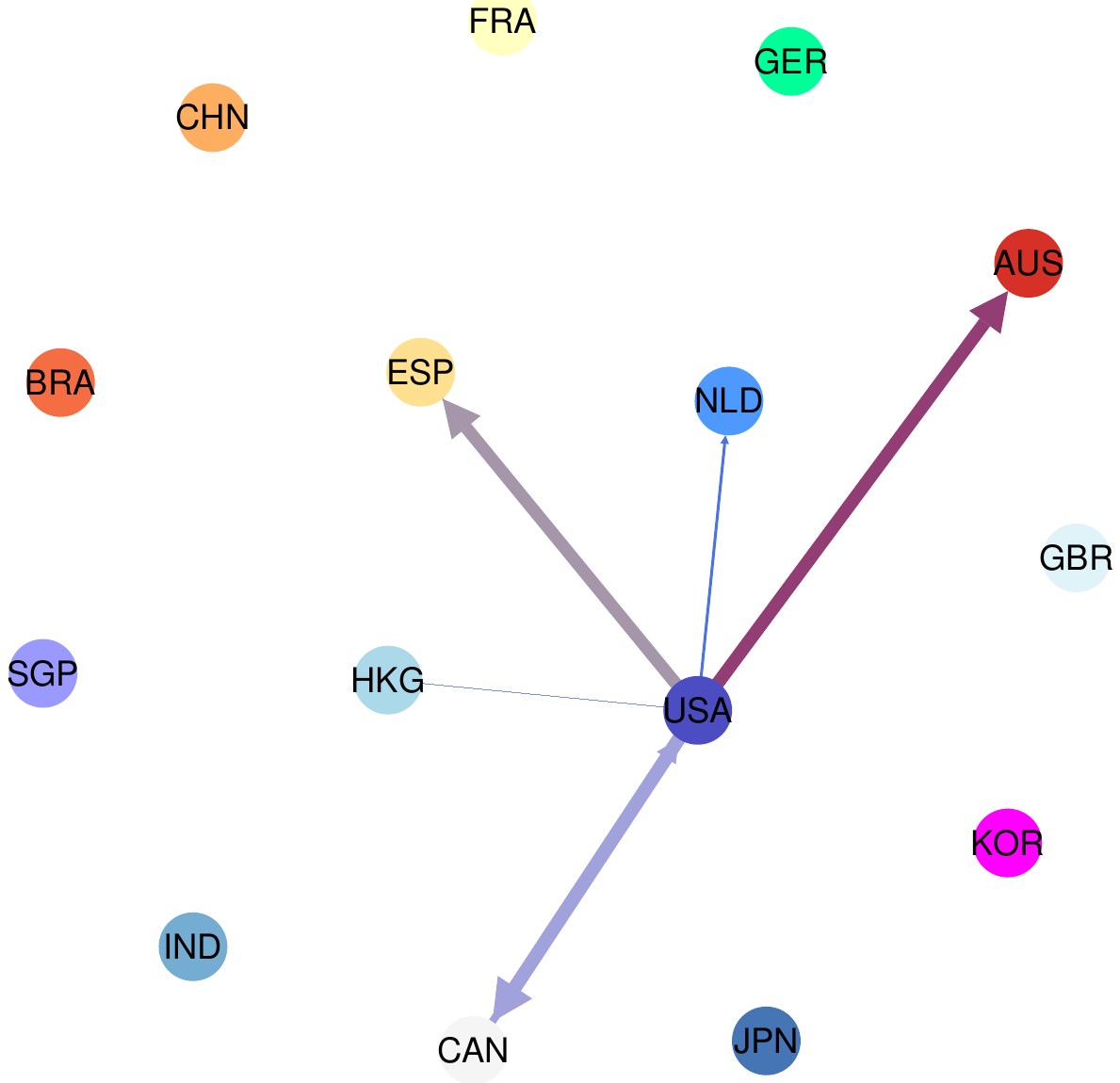}}
  \end{center}
  \caption{ Dynamics of Country Dependency Networks.  We show four networks that are representative for the dynamics within the 13 time windows. The weighted links represent an average p-value of 0.25 or better. This low threshold is necessary since the heterogeneity on the stock-stock interdependency level is large. This indicates a noticeable difference between a comovement analysis on the basis of market indices versus single stocks. It also shows that a large part of this inhomogeneity is captured by the sector-wise grouping.}\label{fig:ccmaps}
\end{figure}

In Figure \ref{fig:ccmaps} we present the dynamics of the country-level dependency networks. We present four networks that are representative for the dynamics within the total of 13 time windows. The weighted links represent an average p-value equal or smaller than 0.25. This low threshold is necessary since the heterogeneity on the stock-to-stock interdependency level is large. This indicates that there is a noticeable difference between a comovement analysis on the basis of market indices versus single stocks. It also shows that a large part of this inhomogeneity is captured by the sector-wise grouping.

\newpage
\section{Number of Links, Sector-by-sector}\label{sec:appsec}

%We counted the number of significant links between the sectors in each country and aggregate for all 13 time windows. Next, the sectors are sorted by the total number of links from top to bottom (and/or left to right). A core of very connected sectors is observed, namely the energy sector, materials sector, and to a slightly lesser extent, the financial sector. The underlying stocks, however, do not only comove with stocks from the same sector, they often also comove with stocks from related sectors. This core is emanating comovement onto other sectors.

\begin{figure}[htb]
\center
\includegraphics[width=0.7\textwidth]{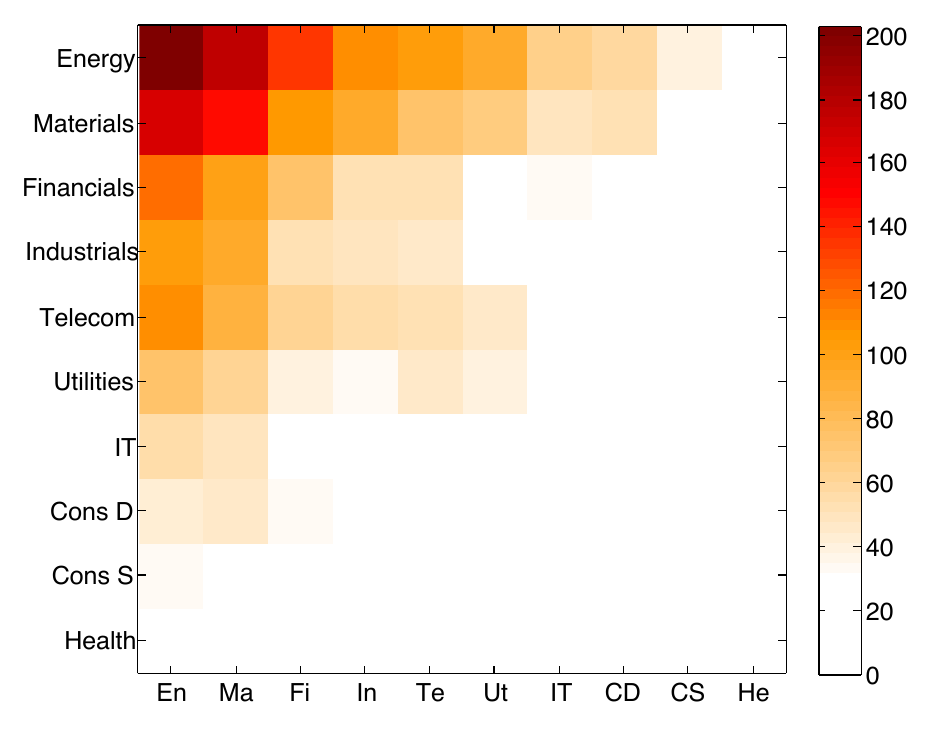}
\caption{\label{fig:secsec} Number of Significant Links for all Time Windows, Sector to Sector, Sorted. The figure shows a color-coded count of the number of links between all combinations of sectors, regardless of the country. The sector names on the horizontal axis are the same as on the vertical in an abbreviated form. The row and column number of each sector has been arranged such that the sum over rows (or columns) is descending. We observe that we do not have significant structures (cliques) composed of combinations of specific sectors, but that the distribution of links on the sector to sector level is roughly the product of the distributions of links on the sector level.}
\end{figure}

%\bigskip
%\centerline{FIGURE \ref{fig:secsec} HERE}
%\bigskip

% It should be noted that if one would only look at links between different countries this impression would be change.



\end{document}